\documentclass{article} 
\usepackage{epsfig}
\usepackage{latexsym}
\usepackage{amsmath}
\usepackage{amssymb}%
\usepackage[english]{babel}
\usepackage{epsfig}
\usepackage{eepic}
\usepackage{mathrsfs} 
\usepackage{xcolor}
\usepackage{bm}
\usepackage{pict2e}
\usepackage{geometry}
\newtheorem{thm}{Theorem}[subsection]
\newtheorem{cor}[thm]{Corollary}
\newtheorem{lem}[thm]{Lemma}
\newtheorem{prop}[thm]{Proposition}
\newtheorem{defn}[thm]{Definition}
\newtheorem{rem}[thm]{Remark}
\newtheorem{expl}[thm]{Example}

\newcommand{\Real}{\mathbb{R}}

\newcommand{\sss}[1]{{\scriptstyle #1}}

\numberwithin{equation}{section}

\newfont{\sfl}{cmssi12}

\usepackage{pict2e}
\usepackage{geometry}
\begin{document}

\title{ From Utility Functions and Generalized Means to Distance Functions: A Stone–Geary Approach and Related Duality Results\thanks{This paper was presented at the “Symposium autour des travaux de Gérard Debreu,” held on September 17–18, 2024. The authors are grateful to Stéphane Vigeant and Nicolas Gravel for helpful suggestions, and especially to Alan Kirman for stimulating discussions. This paper is dedicated to the memory of Robert Russell (1938–2023) and Knox Lovell (1942–2025), who made many fundamental contributions to the field of production economics, and in particular to the topic addressed in this paper.}}

\author{Walter Briec \thanks{\small LAMPS-Laboratory of Mathematics and Physics, University of Perpignan, 52 avenue Villeneuve,
66000 Perpignan, France. }   }
\maketitle
 
\date

\maketitle

\begin{abstract}
This article demonstrates how a large number of efficiency measures known in the literature in production economics can be interpreted through the notion of utility function, based on the concept of Stone-Geary utility. Several relationships between these utility functions and distance functions, a commonly used tool in production theory, are established. To achieve these objectives, a generalized mean distance function is introduced, inspired by the Atkinson inequality index, itself derived from the notion of the Aczel mean. It measures the maximum sum of netput expansions required to reach an efficient point. Several duality theorems are established, linking the new distance functions to the profit function. For all feasible production vectors, the results include as special cases most of the dual correspondences previously established in the literature. Finally, a large class of measures is identified for which these duality results can be obtained without requiring convexity. A numerical example is provided.
\end{abstract}


{\bf Keywords:} { Utility Based Distance Function, Generalized Mean, Directional Distance Function, F\"are-Lovell efficiency measure, Duality Theory, Non-Convexity.  }


%


\section{Introduction}
\label{sec:intro}

The theories of consumers and producers are crucial for understanding the issues surrounding general equilibrium in economics. Consumer theory examines agents' preferences and extends into a broader field of economic analysis due to its importance in decision theory, with the utility function playing a key role in representing these preferences.

On the other hand, producer theory, when considered separately from consumer theory, often focuses on productivity indices and the axiomatic properties of efficiency measures. Like consumer theory, producer theory relies heavily on the concepts of duality and distance functions. These ideas have been particularly emphasized in a series of papers by Luenberger \cite{l92a,l92b, l94, l95, l96}, who drew parallels with Maurice Allais's \cite{a89} surplus measures. Despite their interdependence, these theories have largely developed independently. However, distance functions have proven especially useful in evaluating productive efficiency and have led to significant advancements and applications in operations research. The goal of this article is to establish a formal link between utility functions and distance functions. We will demonstrate that various well-known efficiency measures can be formalized through a type of producer utility function inspired by the Stone-Geary utility concept (see respectively \cite{s54} and \cite{g50}).

In production economics, two primary concepts of technical efficiency are distinguished. The first, introduced by Debreu  \cite{d51}  and especially Farrell \cite{f57}, relates to a radial measure of efficiency. This measure is defined as the minimal proportional reduction of all inputs while maintaining a specific output level and is essentially the inverse of Shepard's distance function \cite{s74}. This foundational work has been expanded to include various non-radial and directional efficiency measures.

The second concept originates from Koopmans' work \cite{k51}, which defines technical efficiency based on the efficient subset of technology. According to this view, a producer is technically efficient if any increase in output or decrease in input requires a corresponding decrease in another output or an increase in another input. This creates potential conflicts between the two efficiency concepts when the isoquant and the efficient subset diverge.

These concepts have been widely applied in economics and studied from the perspective of operations research, leading to further theoretical and axiomatic developments. Färe and Lovell \cite{fl78} introduced a new input efficiency measure, known as the Russell efficiency measure, which averages its component measures. Russell  \cite{r85, r90}  later analyzed these measures, examining their compatibility with various axioms proposed in the literature. The Färe-Lovell measure addresses some limitations of the Debreu-Farrell measure, which evaluates technical efficiency relative to an isoquant rather than an efficient subset. This can result in the misidentification of technical efficiency and produce erroneous inferences. However, Russell noted some continuity issues with the Färe-Lovell measure.

 Subsequent research has proposed various extensions and new measures to overcome the limitations of existing approaches. Luenberger \cite{l92b} introduced the benefit function as a directional representation of preferences, generalizing Shepard's   input distance function \cite{s70}. He demonstrated that Pareto efficiency could be transformed into an optimality principle, the zero-maximum principle. Chambers, Chung, and Färe \cite{ccf96,ccf98} later relabeled this as the directional distance function, which is now a standard in the field. Recent work by Russell and Schorm \cite{rs18} classifies efficiency measures into slacks-based and path-based indices, highlighting the trade-offs between characterizing efficient points and maintaining continuity. More recently, Chambers and Miller \cite{cm14}) contribute to the literature on efficiency measurement by introducing an ordinal model that focuses on comparative, rather than cardinal or absolute, notions of efficiency. They show that these axioms characterize a family of path-based measures related to the coefficient of resource utilisation.

This paper extends these contributions by introducing a broad class of technical efficiency measures based on utility functions, a novel approach in the literature. Inspired by the Stone-Geary utility function, we construct an efficiency measure with theoretically interesting properties. The Stone-Geary utility function originates in comment made by
Geary \cite{g50} on an earlier paper by Klein and Rubin \cite{kr48}, the scope of which was to determine
an appropriate price index in the presence of rationing.{Since then, the Stone--Geary utility function has been widely used in the economic literature, particularly in the analysis of welfare and consumer demand \cite{dm80, dmr82, l09}, as well as in environmental economics \cite{jvdp19, km16}.}
 We define a generalized mean directional distance function, which includes the previously proposed directional distance functions as special cases. This new measure maximizes the generalized sum of netput quantities until an efficient point is reached, drawing on Atkinson's inequality index.{We thus introduce a broad class of efficiency measures and show that many of them, although differing in their formal representation, are in fact linked to a common principle, which can be expressed through a Stone–Geary utility function.}
 
{This extension applies to a wide range of production technologies, including the nonparametric production frameworks introduced by \cite{ccr78} and \cite{bcc84}. We also highlight links between our approach and several strands of the economic literature related to decision-making and multicriteria analysis. In particular, the proposed framework can be connected to bargaining theory (see \cite{ht02, ht10}). Furthermore, it is closely related to goal programming approaches in which the unit used to assess efficiency may be interpreted as an anti-ideal point (see, for example,  \cite{z74,   rtj98} and \cite{g88}).
}

We also derive a general duality result connecting the generalized directional distance function to the profit function, showing that these duality results can be obtained without assuming convexity. Using the concept of the greatest distance from a point to a set (see for instance
\cite{l75} and \cite{nnt23}), we link strong efficiency to norm maximization problems, extending our results to generalized means. This dual correspondence encompasses established results as special cases, providing a comprehensive framework for existing distance functions and efficiency measures. A numerical example in the appendix illustrates our findings.

{The paper is organized as follows. Section~2 is devoted to primal approaches. Section~2.1 introduces the basic framework and reviews key axiomatic concepts in efficiency analysis. Section~2.2 defines a broad class of efficiency measures based on the Stone--Geary utility function and discusses links with decision theory and multicriteria analysis. Section~2.3 examines directional properties using the Atkinson generalized mean and shows how the resulting class of measures relates to well-known efficiency indicators, including the directional distance function and the Färe--Lovell measure. Axiomatic properties are further analyzed, and Section~2.4 focuses on limiting cases, providing a ranking of distance functions with respect to the parameter $p$ of the generalized mean.  
Section~3 addresses dual approaches. Section~3.1 considers the maximization of a norm and shows how duality results from the farthest-distance problem can be used to derive dual representations without relying on convexity assumptions. Section~3.2 extends the analysis to quasi-concave utility functions, while Section~3.3 examines limiting cases and establishes a broad class of duality results when utility functions are weighted by the inverse of the direction. The paper concludes with a taxonomy showing that many existing duality results can be recovered as limiting cases of our framework.}

\section{Utility Based Distance Function and Generalized Means}
\label{sec:basic}
\subsection{Distance  Function, Efficiency and Production Microeconomics: the Groundwork}

A technology describes all the possibilities of production making it possible to transform input vectors into output vectors. The technology can also be modelled by a set of netput
vectors. 
By convention inputs are negative and output positive. Let us denote $T$ the set of all the 
production possibilities (or technology set). We suppose that there are $m$ inputs and $n$ outputs. Therefore 
$d=m+n$ and we write $z = (z_1,z_2,...,z_d )$ where  $d$ is the number of commodities
and $z_k$ denotes the quantity of commodity $k$ in each netput vector. We also assume that the inputs are indexed from $k=1$ to
$m$ and  the outputs from $k=m+1$ to $d=m+n.$ The production set is therefore  a subset of $\Real_-^m\times \Real^n.$  The set valued map $T: \Real_-^{m}\rightrightarrows \Real^n$  defined as $T(x)=\{y:(x,y)\in T\}$ is called the input correspondence. The set valued map  $T^{-1}:\Real^{n}\rightrightarrows \Real_-^m$   defined as $T^{-1}(y)=\{x:(x,y)\in T\}$ is called the output correspondence.   Therefore, if $z:=(x,y)\in T$, then $y\in T(x)$ and $x\in T^{-1}(y)$.

The following standard conditions are imposed on $T$:\\
T1: $(0,0)\in T$ and $y\not=0\Rightarrow (0,y)\notin T$;\\
T2: For all $z\in T$, $\{u\in T: u\geq z\}$ is bounded;\\
T3: $T$ is a closed set and has a non-empty interior;\\
T4: If $z\in T$  and $z'\leq z$ then  $ z'\in T$.\\
T5: $T$ is convex.\\
\noindent Apart  from  the  traditional regularity assumptions (possibility of
inaction, boundedness, and closedness), assumption T4  represents
the strong or free disposability of netputs.{In what follows, for any $z \in T$, we define
$T_z = \{ u \in T : u \geq z \} = T \cap (z + \mathbb{R}^d_+).$ The set $T_z$ consists of all feasible netput vectors in $T$ that weakly dominate $z$. Assumption~T2 implies that $T_z$ is bounded.}

When discussing  efficiency measures and distance functions, it is important to isolate two subsets
of the production set. First, we can define the  efficient subset and the weak efficient subset respectively as follows:
\begin{equation}\label{WEff}
\mathcal E(T)=\{z\in T:z'\geq z \text{ and }z'\not= z\Rightarrow z'\notin T\}\quad \text{and}\quad \mathcal F (T) = \{z\in T:z'> z \Rightarrow z'\notin T\}.
\end{equation}

Obviously, these two subsets of the production set are connected via the following inclusions:
$\mathcal E (T) \subseteq \mathcal F (T) \subseteq T$.{Note that the efficient subset is also referred to as the \emph{strictly efficient set} in \cite{ht02} and can be written as
$\mathcal{E}(T) = \{ z \in T : T_z \cap T = \{ z \} \}.$ }

Now, recall the definition of the directional distance function defined in
\cite{ccf98}. The directional distance function is the map
$D_T:\Real^d_+\times \Real^d_+ \longrightarrow
\Real_+\cup\{-\infty,\infty\}$ defined by:
\begin{align}\label{DF}
\mathrm{D}(z;g) =  \sup\limits_{\delta \in \Real}\left\{\delta: z+\delta g\in T\right\}.
\end{align}
If the condition is not vacuous, that is there is a $\delta$ with $z+\delta g\in T$ and $g\not=0$, then the sup will be achieved; otherwise it is defined to be $-\infty$. If $z\in T$ and $g=0$, then $D_T(z,0)=+\infty$.  
In words, this directional distance function indicates the maximal expansion in the direction of $g$ simultaneously in all netputs which still allow the production.

A well-known property of the   directional
distance function is that it exhibits the translation invariance property. Under a strong disposability assumption with $g\in \Real_{++}^d$,
$D (z;g)=0\Leftrightarrow z\in \mathcal F (T)$. However, in the general case, $D (z;g)=0$    does not
imply that $z$ belongs to $\mathcal E (T)$.   Hence,
the computation of $\mathrm{D}(z;g)$ for a given direction $g\in \Real_+^d$ does not
allow to conclude whether a production plan is efficient or not.
  In fact, $z\in \mathcal E (T)$
implies that $\mathrm{D}(z;g)=0$, but the converse is only true in the
special case when $\mathcal E (T)= \mathcal F(T)$ and $g\in \Real_{++}^d$.  For all $g\in \Real_+^d$ the support of $g$ is defined as $    \mathcal G=\{k\in [d]: g_k\not=0\}$. This implies that $ \mathcal G=\{k\in [d]:g_k>0\}$.  

 Along with this approach, one can extend to the full netput space the asymmetric directional distance function defined by \cite{bck11} as
$\mathrm{AD}:\Real^d\times\Real^d_+\longrightarrow\Real_+\cup\{-\infty, \infty\}$,
\begin{align}\label{AF}
\mathrm{AD}(z;g) = \max\limits_{k\in \mathcal G}\mathrm{D}(z;g_ke_k)
\end{align}
where $\{e_k\}_{k\in [d]}$ stands for the canonical basis of $\Real^{d}$. This asymmetric directional distance function takes the maximum of the dimension-wise reduction in each netput direction, which allow production.{A closely related approach is proposed in \cite{ht02}, where the authors consider the maximal value of each coordinate in the set $T_z$. In this spirit, an ideal production vector relative to $z$ can be defined as the vector whose components correspond to the projections of $z$ along each canonical direction. In our framework, this ideal production vector is given by $m(z; g) = \sum_{k \in \mathcal G} \mathrm{D}(z; g_k e_k)\, e_k.$ This construction is closely related to compromise programming in multicriteria decision theory (see \cite{z73, z74}). It also plays a central role in bargaining theory, where it is used to characterize the Kalai--Smorodinsky solution, which is axiomatized by Pareto optimality, symmetry, affine invariance, and restricted monotonicity (see \cite{ht03}).}

\begin{center}
{\scriptsize

\unitlength 0.4mm 
\linethickness{0.4pt}
\ifx\plotpoint\undefined\newsavebox{\plotpoint}\fi 


\begin{picture}(302.25,157)(0,0)
\put(130,35.75){\vector(-1,0){121.75}}
\put(129.75,36.25){\vector(0,1){110.5}}
\put(47.5,79){\circle*{1.414}}
\put(145.25,55.5){\vector(3,4){.07}}\multiput(129.75,36)(.0336956522,.0423913043){460}{\line(0,1){.0423913043}}
\put(96.5,65.25){\makebox(0,0)[]{$T$}}
\put(134.75,31.25){\makebox(0,0)[]{$0$}}
\put(131.5,157){\makebox(0,0)[]{$y$}}
\put(47.75,75.25){\makebox(0,0)[cc]{$z$}}
\put(99.5,123.5){\makebox(0,0)[cc]{$z^\star$}}
\put(81.25,114.75){\circle*{1.581}}
\put(47.25,1.25){\makebox(0,0)[cc]{{\bf Figure 1:}  Directional Distance Function.}}
\put(152.25,62.5){\makebox(0,0)[cc]{$g$}}
\put(156,35.25){\vector(-1,0){.07}}\multiput(280.5,35)(-15.5625,.03125){8}{\line(-1,0){15.5625}}
\put(280.5,35.25){\vector(0,1){108.5}}
\put(181.25,78.5){\circle*{1}}
\put(81.75,116){\vector(1,1){.07}}\multiput(47.5,79.25)(.03371062992,.03617125984){1016}{\line(0,1){.03617125984}}
\put(238.5,64.25){\makebox(0,0)[]{$T$}}
\put(284.75,32){\makebox(0,0)[]{$0$}}
\put(280.5,151.5){\makebox(0,0)[]{$y$}}
\put(181.75,75.5){\makebox(0,0)[cc]{$z$}}
\put(261.75,79){\circle*{1.581}}
\put(181.75,139){\circle*{1.581}}
\put(209,0){\makebox(0,0)[cc]
{{\bf Figure 2:} Asymetric Directional Distance Function}}
\put(0,35.5){\makebox(0,0)[cc]{$x$}}
\put(149,35){\makebox(0,0)[cc]{$x$}}
\put(129.75,43.25){\line(0,-1){34.75}}
\put(280.5,42.5){\line(0,-1){34.75}}
\put(280.75,35){\vector(1,0){13.75}}
\put(280.5,34.75){\vector(0,1){13.75}}
\put(302.25,34.5){\makebox(0,0)[cc]{$e_1$}}
\put(287.25,50.75){\makebox(0,0)[cc]{$e_2$}}
\put(181,78.75){\vector(0,1){60}}
\put(180.75,78.75){\vector(1,0){81}}
\qbezier(90.25,95)(77.375,141)(17,144)
\qbezier(130,35.5)(98.25,67.5)(90.5,94.5)
\qbezier(280.75,35.25)(273.625,87.125)(235,89.5)
\qbezier(234.75,89.75)(197.875,92.5)(176.5,153.25)
\end{picture}

}
\end{center}

By construction it is implicitly assumed that $z\in T$ to ensure a feasibility condition. This is a useful assumption to establish the main results of the paper. In the following, for all finite subsets $A$, $|A|$   stand for the cardinality of $A$. 

The F\"are-Lovell efficiency measure \cite{fl78} emerged from a debate on axiomatic properties of radial efficiency
measures. This
function $\mathrm{E}_{\mathrm{FL}}:\Real^m_-\backslash\{0\}\times\Real_+^n
\longrightarrow\Real_+\cup\{\infty\}$ is defined for all $z:=(x,y)\in T$ as
follows:
\begin{align}\label{FL}
\mathrm{E}_{\mathrm{FL}}(x,y) =\inf\limits_{\beta \in
[0,1]^{m}}\Big\{\frac{1}{|\sss {\mathcal I( x)}|} \sum\limits_{i\in \sss {\mathcal I(x)}}
\beta_i: (\beta \odot x ,y  )\in T \Big\},
\end{align}
\noindent If $z\notin T$  it is defined to be $+\infty$. The symbol $\odot$ denotes the Hadamard product (element by element)\footnote{For all $w,z\in \Real^d$, $w\odot z=(w_1z_1,....,w_dz_d)$.}
of two vectors, and for all $x\in \Real^n$ the support
of $x$ is defined as $  \small {\mathcal I(x)}=\{i: x_i\not=0\}$ and its cardinal is $|  \small X|$. This F\"are-Lovell efficiency measure 
 indicates the minimum average sum of
dimension-wise reductions in each input dimension which maintains
production of given outputs on the efficient subset of the input
set.

Since the directional distance function fails to characterize the efficient subset, one can
 define a  F\"are-Lovell   directional distance function as
follows. The  F\"are-Lovell directional distance function is the map
$\mathrm{D}_{\mathrm{FL}}:\Real^d\times\Real^d_+\longrightarrow \Real_+\ \cup\{-\infty,+\infty\}$
 defined for all $z\in T$ as:
\begin{align}\label{DFL}
\mathrm{D}_{\mathrm{FL}}(z;g) = \sup \limits_{\delta \in \Real_+^{d}}
\Big\{ \frac{1}{ |\scriptstyle{{\mathcal G}}|   }\sum\limits_{k\in \mathcal  G}\delta_k :z+\delta \odot g\in T\Big\}.
\end{align}
If $z\notin T$ then it is defined to be $-\infty$. If $z\in T$ and $g=0$, then $\mathrm{D}_{\mathrm{FL}}(z;g) =+\infty$.
{This approach was introduced in \cite{fw09} in a nonparametric setting. It is also closely related to the slack-based measure proposed in \cite{t01} and extends the definition put forward in \cite{bck11} along two dimensions.
} First it proposes a measure oriented in the full netput space. Second it takes into account the cardinality of the set $ \mathcal G$. This is important for some of the results further established in the paper.  In particular if $g=(x,0)$ then:
\begin{equation}
\mathrm{D}_{\mathrm{FL}}(x,y;x,0)=1-\mathrm{E}_{\mathrm{FL}}(x,y).
\end{equation}
In addition, note that if $g=0$ then $ \small \mathcal G=\emptyset$ and, by construction, $\mathrm{D}_{\mathrm{FL}}(z;g)=\mathrm{D}_{\mathrm{FL}}(z;0)=-\infty$.

\begin{center}
{\scriptsize

\unitlength 0.4mm 
\linethickness{0.4pt}
\ifx\plotpoint\undefined\newsavebox{\plotpoint}\fi 
\begin{picture}(293.75,157)(0,0)
\put(130,35.75){\vector(-1,0){121.75}}
\put(129.75,36.25){\vector(0,1){110.5}}
\put(51,81){\circle*{1.814}}
\put(145.25,55.5){\vector(3,4){.07}}\multiput(129.75,36)(.0336956522,.0423913043){460}{\line(0,1){.0423913043}}
\put(96.5,65.25){\makebox(0,0)[]{$T$}}
\put(134.75,31.25){\makebox(0,0)[]{$0$}}
\put(131.5,157){\makebox(0,0)[]{$z_2$}}
\put(51.25,77.25){\makebox(0,0)[cc]{$z$}}
\put(99.5,123.5){\makebox(0,0)[cc]{$z^\star$}}
\put(84.75,117.75){\circle*{1.814}}
\put(47.25,1.25){\makebox(0,0)[cc]{{\bf Figure 3:}  Directional F\"are-Lovell}}
\put(152.25,62.5){\makebox(0,0)[cc]{$g$}}
\put(163.75,137){\vector(-1,0){.07}}\multiput(288.25,136.75)(-15.5625,.03125){8}{\line(-1,0){15.5625}}
\put(85.25,118){\vector(1,1){.07}}\multiput(51,81.25)(.03371062992,.03617125984){1016}{\line(0,1){.03617125984}}
\put(209,0){\makebox(0,0)[cc]{{\bf Figure 4:} F\"are-Lovell Measure}}
\put(0,35.5){\makebox(0,0)[cc]{$z_1$}}
\put(129.75,43.25){\line(0,-1){34.75}}
\put(283.5,18){\vector(0,-1){.07}}\multiput(284,142.25)(-.0333333,-8.2833333){15}{\line(0,-1){8.2833333}}
\put(152.75,124){\line(1,0){97.25}}
\qbezier(250,124)(269.75,123.875)(274.5,21.25)
\put(167,112.75){\circle*{2.062}}
\multiput(166.93,112.93)(.966942,.200413){122}{{\rule{.4pt}{.4pt}}}
\put(250.25,124.25){\circle*{2}}
\put(168.75,98.25){\makebox(0,0)[cc]{$x$}}
\put(264.25,123.25){\makebox(0,0)[cc]{$x^\star$}}
\put(215,69){\makebox(0,0)[cc]{$T^{-1}(y)$}}
\multiput(7.93,115.93)(-.125,.75){3}{{\rule{.4pt}{.4pt}}}
\qbezier(95.25,118)(121.625,118)(129.5,36)
\put(95.5,117.75){\circle*{1.5}}
\put(79.75,123.5){\makebox(0,0)[cc]{$z'$}}
\put(243.25,130.25){\makebox(0,0)[cc]{$x'$}}
\put(285,10){\makebox(0,0)[cc]{$x_2$}}
\put(155,135){\makebox(0,0)[cc]{$x_1$}}
\put(293.75,143.75){\makebox(0,0)[cc]{$0$}}
\put(95,118){\line(-1,0){29.25}}
\qbezier(66.25,118.25)(36.875,118.125)(9,134.5)
\end{picture}

}
\end{center}

In Figure 3, $z'$ is projected  from $z$ in the direction of $g$. It is a weakly but not (strongly) efficient point. $z'$ should be pushed to $z^\star$ to reach the efficient subset $\mathcal E(T)$. In Figure 4, $x'$ is not an efficient input, while $x^\star$ is efficient. It follows that the value of the F\"are-Lovell measure at point $x^\star$ is equal to 1.

 



\subsection{A Utility-Based Formalism for Distance Functions and Efficiency Measures}\label{UForm}
We propose a novel representation of the concept of a distance function. Later, we will demonstrate that many commonly used distance functions in the efficiency analysis literature can be expressed within this framework.

The proposed approach closely resembles the method used to define the Stone-Geary utility function. Generally, the Stone-Geary utility function is derived from the Cobb-Douglas utility function:
\begin{equation}
W_{{0}}(z_1, \dots, z_d) = \prod_{k \in [d]} (z_k - \gamma_k)^{\alpha_k},
\end{equation}
where \( z_k \) represents a consumption good, and \( \gamma \) and \( \alpha \) are parameters with \( \alpha \in \mathbb{R}_+^d \) and \( \sum_{k \in [d]} \alpha_k = 1 \).
 The Stone-Geary function is used to model problems involving subsistence levels of consumption.  In these cases, a certain minimal level of some good has to be consumed. In our context a producer seeks to perform better from a minimal level of netputs. For $p\leq 1$ and $p\not=0$ we have a CES formulation involving the function:
\begin{equation}W_{{p}}(z_1,...,z_d) =\stackrel{\phi_p}{\sum\limits_{k\in
[d]}}{\alpha_k}^{\frac{1}{p}}(v_k-\gamma_k).  \end{equation}
We propose a new class of distance functions inspired by this formalism.  To simplify the technical exposition, for all $z\in T$, let
  \begin{equation}
T_{z}=\big \{u\in T: u\geq z\big \}
\end{equation}
denote the set of the netput vectors that dominate $z$. 

\begin{defn}\label{DefUbased} A distance function  $\mathscr D : \Real^d\longrightarrow \Real_+ \cup\{-\infty,+\infty\}$  admits a utility-based representation if there is a map  $W:\Real^d\longrightarrow \Real$, such that $(i)$ $W(0)=0$; $(ii)$ $W$ is upper semi-continuous; $(iii)$ $W$ is non-decreasing; $(iv)$ For all $z\in T$  
\begin{align}
\mathscr D (z )= \sup_{ u\in T_{z}} W(u-z),
\end{align}
where  $\mathscr D$ is defined to be $-\infty$ when $z\notin T$.

\end{defn}
Note that by hypothesis, since $W$ is nondecreasing, if $u\geq 0$, then $W(u)\geq 0$. In the remainder we say that a distance function satisfying the conditions of Definition \ref{DefUbased} with respect to $W$ is a {\bf $W$-utility based distance function}.  Conversely, note that if $T1$–$T4$ hold, it is always possible to construct a distance function $\mathscr{D}_W$ from a map $W$ satisfying conditions $(i)$, $(ii)$, and $(iii)$ in Definition \ref{DefUbased}.\\
From $T2$ and $T3$, the set $T_z$ is compact for any $z \in T$. Therefore, $\mathscr{D}_W$ is well-defined, and there exists some $u^\star \in T_z$ such that $W(u^\star - z) = \mathscr{D}_W(z)$.

The basic intuition is that distance functions of this class aim to capture the maximum improvement from a point $z$ to a boundary point $z^\star$. This improvement is measured by a function that resembles a utility function, constructed following the Stone-Geary utility framework.

{In this respect, the proposed approach is closely related to several issues arising in decision theory, multicriteria analysis, and bargaining theory. In the context of goal programming, the feasible point $z$ whose efficiency is evaluated plays a role analogous to that of an anti-ideal point (see, for instance, \cite{z74, rtj98}). In multiobjective optimization, the anti-ideal point (also referred to as the nadir point) is defined as the vector whose components correspond to the worst attainable values of each objective over the Pareto-efficient set. It represents the least desirable performance level that can be simultaneously achieved across all criteria within the feasible region.  The present approach follows a similar logic, though restricted to the subset $T_z$. The function $\mathscr{D}_W$ measures the maximal attainable improvements by maximizing a distance-based function between the point $z$ and the production frontier.  A related analogy can be drawn with the bargaining framework, in which $z$ plays the role of the disagreement point associated with the feasible set $T_z$. Classical solutions to the bargaining problem include the Nash solution \cite{n50}, based on a multiplicative criterion, and the Kalai--Smorodinsky solution \cite{ks75}. Hougaard and Tvede establish explicit connections between bargaining theory and production economics in \cite{ht02} and \cite{ht10} (see also \cite{ht03} for related results).  
It should be noted, however, that the problem of efficiency measurement differs in some respects from those typically addressed in goal programming or bargaining frameworks. In the present setting, the primary objective is to compare production vectors within a given technology, while benchmarking considerations associated with specific efficiency measures are only treated implicitly. Nonetheless, Hougaard and Tvede have shown that such issues can be fruitfully analyzed using tools and concepts drawn from bargaining theory. Finally, note that there also exist studies showing that the Cobb–Douglas–Stone–Geary utility function can be used in a bargaining context (see, for instance, \cite{l09}).}

{Note also, that this construction is closely related to the notion of \emph{sup–convolution}, the counterpart of the inf–convolution introduced by Moreau \cite{m65}. For functions $f, g : \mathbb{R}^d \to \mathbb{R} \cup \{-\infty\}$, the sup–convolution is defined as
\[
(f \Box g)(z) = \sup_{u \in \mathbb{R}^d} \{ f(u) + g(z-u) \}.
\]
Let $\bm{\delta}_E$ denote the indicator function of a set $E \subset \mathbb{R}^d$, with $\bm{\delta}_E(u) = 0$ if $u \in E$ and $-\infty$ otherwise. Then, for any $z \in T$, $\mathscr{D}_W(z)$ can be written as
\[
\mathscr{D}_W(z) = \sup_{u \in T_z} W(u-z) 
= \sup_{u \in \mathbb{R}^d} \{ \bm{\delta}_{T_z}(u) + W(u-z) \}.
\]}

 In the following, we say that $W$ is {\bf strongly increasing } (strongly decreasing) if $z\not=z'$ and $z'\geq z$   implies that 
$W(z')>W(z)$ ($W(z')<W(z)$).  $W$ is {\bf weakly increasing } (weakly decreasing) if $z>z'$ implies that
$W(z')>W(z)$ ($W(z')<W(z)$).   $W$ satisfies an {\bf absorption condition} if $z_k=0$ for some $k$ implies $W(z)=0$. The generalized mean is an example of applications that satisfy this absorption condition when $p\leq 0$ (including the multiplicative case). 

\begin{prop}\label{UtForm}Suppose that $\mathscr D_W  :\Real^d\longrightarrow \Real_+\cup\{-\infty \}$ is  a  $W$-utility-based distance function, where  $W:\Real^d\longrightarrow \Real$, such that $(i)$ $W(0)=0$; $(ii)$ $W$ is upper semi-continuous ; $(iii)$ $W$ is non-decreasing. Under $T1$ to $T4$,   $\mathscr D_W $ satisfies the following properties:\\
$(a)$ If $W$ is strongly increasing then $\mathscr D_W $ is strongly decreasing. Moreover $\mathscr D_W (z)=0$ if and only if $z\in \mathcal E(T)$. \\
$(b)$ If $W$ is weakly increasing then $\mathscr D_W $ is weakly decreasing. Moreover if $\mathscr D_W $ satisfies the absorption condition then $\mathscr D_W (z)=0$ if and only if $z\in \mathcal F(T)$.\\
$(c)$ If $W$ is quasi-concave and   $T$ is convex then $\mathscr D_W $ is quasi-concave  over $T$. \\
$(d)$ If $W$ is upper semi-continuous then $\mathscr D_W $ is upper semi-continuous over $T$. \\

\end{prop}

In the following we propose an example in the case of a non-parametric technology (see \cite{bcc84}). A related numerical example is proposed in appendix. 

\begin{expl}\label{NonPar}Let $A\subset \Real_-^m\times \Real^n$ be a finite set of observed production vectors. Following \cite{bcc84}, one can define the production technology:
$$T_A=\big\{u\in \Real^d: u\leq \sum_{a\in A}t_a a:\sum_{a\in A}t_a=1, t\geq 0  \}.$$ It follows that we have:

$$\mathscr D_{W}(z)= \sup_{  \delta\geq 0,  t\in \Real_+^{|A|} } \Big \{W(\delta ):\delta = \sum_{a\in A}t_aa -z, \sum_{a\in A}t_a=1,  t\geq 0 \Big\}.$$
Notice that if $W$ is concave then the problem can be solved by concave programming methods. 
\end{expl}

The following results are useful in the remainder of the paper in which we   consider some specific sequences of functions based on the notion of generalized mean.

\begin{prop}\label{LimGen} Let $\{\mathscr  D_{W_{{p}}}\}_{p\in \mathbb N}$ be a sequence of $W_{{p}}$-utility based distance functions where $\{W_{{p}}\}_{p\in \mathbb N}$ is a  sequence of functions satisfying   conditions $(i)$, $(ii)$ and $(iii)$ in definition 
\ref{DefUbased}. Suppose moreover that, for each $p$, $W_{{p}}$ is continuous and for any compact subset $ K$ of $\Real_+^d$ the sequence   $\{W_{{p}}\}_{p\in \mathbb N}$ uniformly converges to a function  $W$. Then the map $\mathscr D_W : \Real^d\longrightarrow \Real_+\cup\{-\infty\}$ defined for all $z\in T$ as:
\begin{align}
\mathscr D_W(z )= \sup_{ u\in T_{z}} W(u-z) 
\end{align}
is a $W$-utility-based distance function where  $\mathscr D_W$ is defined to be $-\infty$ when $z\notin T$. Moreover, for all $z\in T$
$$\lim_{p\longrightarrow \infty}\mathscr D_{W_{{p}}}(z)=\mathscr D_W(z). $$
\end{prop}

The following statement uses the fact that if a monotone sequence of continuous functions pointwise converges to a continuous function on a compact set, then it converges uniformly.
 
\begin{cor}\label{CorLimGen} Let $\{\mathscr D_{W_{{p}}}\}_{p\in \mathbb N}$ be a sequence of $W_{{p}}$-utility based distance functions where $\{W_{{p}}\}_{p\in \mathbb N}$ is a  monotonic sequence of functions satisfying   conditions $(i)$, $(ii)$ and $(iii)$ in definition 
\ref{DefUbased}. Suppose moreover that for each $p$ $W_{{p}}$ is continuous and for any compact subset $ K$ of $\Real_+^d$ the sequence   $\{W_{{p}}\}_{p\in \mathbb N}$   is pointwise convergent to a continuous function  $W$ over   $K$. Then the sequence $\{\mathscr  D_{W_{{p}}}\}_{p\in \mathbb N}$   converges to the   map $\mathscr D_W : \Real^d\longrightarrow \Real_+\cup\{-\infty\}$ defined for all $z\in T$ as:
\begin{align}
\mathscr D_W (z )= \sup_{ u\in T_{z}} W(u-z) 
\end{align}
where  $\mathscr D_W$ is defined to be $-\infty$ when $z\notin T$.
\end{cor}

  {In the following, a suitable notion of generalized mean is considered.  For all $p\in ]0,+\infty[ $, let $\phi_p:\Real_+\longrightarrow
\Real$ be the map defined by
$\phi_{p}(\lambda) =
 {\lambda}^{{p}}$.
For all $p\not=0$, the reciprocal map is
$\phi_{{p}}^{-1}:=\phi_{\frac{1}{p}}$. First, it is
quite straightforward to state that:
 $(i)$ $\phi_{p}$ is defined over $\mathbb{R}_+$;
 $(ii)$ $\phi_{p}$ is continuous over $\mathbb{R}_+$; and
 $(iii)$ $\phi_{p}$ is bijective over $\mathbb{R}_+$.
Second, let us focus on the  case $p\in ]-\infty,0[$. The map
$x\mapsto x^{{p}}$ is not defined at point $x=0$. Thus, it
is not possible to construct a bijective endomorphism on
 $\Real_+$.}

 {For all $p\in ]-\infty,0[$ we consider the function $\phi_p:\Real \longrightarrow \Real_{++}\cup \{\infty\}$ and its reciprocal ${ \phi}_{{\frac{1}{p}}}$ respectively defined by:}
\begin{equation}\label{powerneg} {
{ \phi}_{p}(\lambda) =\left\{\begin{array}{ll}
\lambda^{p}&\hbox{ if } \lambda > 0\\
  +\infty&\hbox{ if } \lambda =0 .   \end{array}
                \right.}
\quad \text{ and }\quad 
{ \phi}_{{\frac{1}{p}}}(\lambda) =\left\{\begin{array}{ll}
\lambda^{\frac{1}{p}}&\hbox{ if } \lambda > 0\\
  0&\hbox{ if } \lambda =\infty .   \end{array}
                \right.
\end{equation} 

 {Let us investigate the $\phi_p$ generalized sum analyzed in \cite{b77}. First one can introduce the binary operation $\stackrel{p}{+}$ defined for all $s,t\in \phi_p^{-1}(\Real)$ as
 \begin{equation}
 s\stackrel{p}{+}t= \phi_p^{-1}\Big(  \phi_p(s)+\phi_p(s)\Big).
 \end{equation}
 This binary operation can be extended to a suitable generalized sum defined as follows.
For all $(\delta_1,\ldots,\delta_d) \in \Real_{+}^{d}$, and for all $p>0$ the $\phi_p$-generalized sum is given by:
\begin{equation}
\stackrel{\phi_p}{\sum_{k\in [d]}} \delta_k := \phi_p^{-1}\Big({\sum_{k\in [d]}} \phi_p(\delta_k)\Big) =
 \Big( \sum_{k\in  [d]} \left(\delta_k\right)^p\Big)^{\frac{1}{p}}.
\end{equation}}

 { If $p <0$, using the symbolism $\frac{1}{0}=+\infty$ we have, by construction: }

\begin{equation} {
\stackrel{\phi_p}{\sum_{k\in [d]}} \delta_k := \phi_p^{-1}\Big({\sum_{k\in [d]}} \phi_p(\delta_k)\Big)  =\left\{
\begin{array}{ll}
 \left( \sum_{k\in [d]} \left(\delta_k\right)^p\right)^{\frac{1}{p}}  &\text{ if }\min_k \delta_k>0 \\
0 &  \text{ if
 }\min_k\delta_k=0.
\end{array}
\right.  }
\end{equation}
 Briec, Dumas, and Mecki \cite{bdm19} consider these generalized sums to propose a distance function aimed at measuring efficiency in a social welfare context. Atkinson \cite{a70} also uses a similar formulation to introduce a broad class of social welfare functionals. It is worth noting that Chavas and Cox \cite{cc99} proposed a parameterized extension of the Debreu-Farrell measure using a graph-based approach related to the directional distance function. However, their construction did not involve the notion of generalized means and did not characterize strong efficiency.

It is important to note that we have the following limit properties:

\begin{enumerate}  \item[(i)] $\lim_{p\longrightarrow -\infty}\big( \frac{1}{d}\sum\limits_{k\in
[d]} \delta_k^p   \big)^{1/p}=\min \limits_{k\in
[d]} \delta_k^{\frac{1}{d}} $;\\

\item [(ii)]   $\lim_{p\longrightarrow 0^-}\big( \frac{1}{[d]}\sum\limits_{k\in
[d]} \delta_k^p   \big)^{1/p}=\prod \limits_{k\in
[d]} \delta_k^{\frac{1}{d}} ;$ \\
 
\item [(iii)]     $\lim_{p\longrightarrow +\infty}\big( \frac{1}{d}\sum\limits_{k\in
[d]} \delta_k^p   \big)^{1/p}=\max \limits_{k\in
[d]} \delta_k^{\frac{1}{d}} .$  \\
  
   \end{enumerate}

The geometric structure involved by this class of generalized means is depicted in Figure 6.
\begin{center}{\scriptsize  
\unitlength 0.4mm 
\linethickness{0.4pt}
\ifx\plotpoint\undefined\newsavebox{\plotpoint}\fi 
\begin{picture}(134.75,158.25)(0,0)
\put(130,34.5){\vector(-1,0){121.75}}
\put(129.75,35){\vector(0,1){120.5}}
\put(51.5,83.5){\circle*{1.414}}
\put(51.5,83.5){\line(0,1){60.5}}
\put(51.5,83.5){\line(1,0){80.5}}
\put(96.5,64){\makebox(0,0)[]{$T$}}
\put(76.5,94){\makebox(0,0)[cc]{$T_{z}$}}
\put(134.75,30){\makebox(0,0)[]{$0$}}
\put(131.5,162.75){\makebox(0,0)[]{$y$}}
\put(47.75,74){\makebox(0,0)[cc]{$z$}}
\put(91,107.75){\circle*{1.581}}
\put(96,113.75){\makebox(0,0)[]{$z^\star$}}
\put(47.25,11){\makebox(0,0)[cc]
{{\bf Figure 5:} $W$-utility based Distance Function}}
\put(0,34.25){\makebox(0,0)[cc]{$x$}}
\put(129.75,42){\line(0,-1){34.75}}
\qbezier(36,136.25)(116,112.5)(130,34.75)
\qbezier(4.75,156)(17,142.25)(35.25,136.5)
\put(20.0,116){\makebox(0,0)[cc]{$W(0)$}}
\put(80,153.5){\makebox(0,0)[cc]{$W(z^\star-z)$}}
\qbezier(46.25,158.25)(66.375,154.75)(72,139.25)
\qbezier(124.5,76.75)(118.25,98.125)(109,99)
\qbezier(72,139.25)(85.625,102.25)(108.75,99.25)
\qbezier(14.5,129)(34.625,121)(43.25,93)
\qbezier(43.25,93)(46.625,82.5)(75.5,72)
\qbezier(75.75,71.75)(82.5,68.875)(91.25,55.5)
\end{picture}
\scriptsize
\unitlength 0.33mm 
\begin{picture}(254,209.25)(0,10)
\put(69.25,56.75){\vector(0,1){142.75}}
\put(69,56.5){\vector(1,0){170.5}}
\put(139.5,124){\circle*{1.5}}
\put(139,124.5){\line(-1,0){98.5}}
\put(139.5,123.75){\line(0,-1){86.75}}
\multiput(69.25,189.5)(.036504945831,-.033737635422){4246}{\line(1,0){.036504945831}}
\put(139.75,124.25){\line(1,0){99.25}}
\put(149.75,132.25){\makebox(0,0)[cc]{$(1,1)$}}
\put(139.5,124.25){\line(0,1){78.5}}
\qbezier(102.75,190.75)(125.625,107.75)(206,90.75)
\qbezier(123.25,203)(123.125,142.375)(139.5,125.25)
\qbezier(139.75,124.75)(160.375,108.875)(229.5,102.5)
\qbezier(48.25,141.25)(128.75,137.25)(139.25,124.25)
\qbezier(164.25,43.25)(155.875,109.5)(140,123.75)
\put(248.25,56.25){\makebox(0,0)[cc]{$x_1$}}
\put(67.75,209.25){\makebox(0,0)[cc]{$x_2$}}
\put(63.25,51.5){\makebox(0,0)[cc]{$0$}}
\put(229,43.75){\makebox(0,0)[cc]{$p=1$}}
\put(125.75,41){\makebox(0,0)[cc]{$p=\infty$}}
\put(170.75,40.25){\makebox(0,0)[cc]{$p>0$}}
\put(217,86.5){\makebox(0,0)[cc]{$p=0$}}
\put(239,101){\makebox(0,0)[cc]{$p<0$}}
\put(254,122.75){\makebox(0,0)[cc]{$p=-\infty$}}
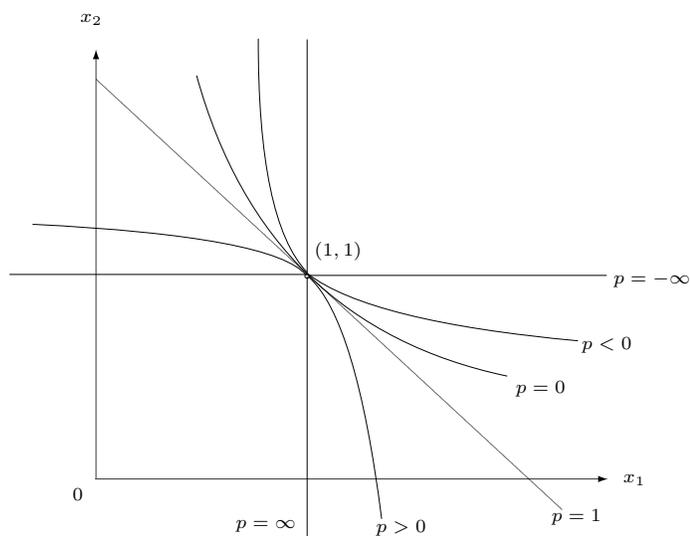
\put(140.75,30){\makebox(0,-5)[cc]{{\bf Figure 6:} Curves $\{(x_1,x_2): x_1\stackrel{p}{+}x_2=2^\frac{1}{p}\}$}}
\end{picture}}
\end{center}

 Briec, Kerstens and Cavaignac { \cite{bck20} show
that, for all $p\in \mathbb R\backslash \{0\}$, this generalized sum is continuous.} {Namely, for all $p\in \mathbb R\backslash \{0\}$, the map $\delta\mapsto \stackrel{\phi_p}{\sum\limits_{k\in [d]}} \delta_k$ is continuous over $\Real_+^d$.}

Now, for any $p \in \Real \cup\{-\infty\}$, we will say that $W_{{p}}$ is a {generalized \bf $p$-mean utility function} if $W_{{p}}$  is defined as follows for each value of $p$:
\begin{enumerate}  \item[(i)] $W_{{p}}(\delta)=\frac{1}{ {d}^{\frac{1}{p}}} \stackrel{\phi_p}{\sum\limits_{k\in
[d]}}\, {\delta_k}  $ if $p\in \Real\backslash\{0\}$;\\

\item [(ii)]   $W_{{0} }(v)=\prod\limits_{k\in [d]} \big( {\delta_k} \big)^{\frac{1}{d}} $ if $p=0$;
 
\item [(iii)]     $W_{-\infty}(\delta)=\min\limits_{k\in [d]}   {\delta_k}  $ if $p=-\infty$; \\
  
 \item [(iv)]    $W_\infty(\delta)=\max\limits_{k\in [d]}  {\delta_k}  $ if $p=\infty$.\\
  \end{enumerate} 
Notice that for each $p$, $W_{{p}}$ is a continuous and non-decreasing map. Therefore, for any $p$, one can construct a $W_p$-utility-based distance function as:

\begin{equation}
\mathscr D_{W_{{p}}}(z)=\sup_{u\in T_z}W_{{p}}(u-z). 
\end{equation}
In particular, if $p\leq 0$ then $W_{{p}}$ satisfies the absorption condition, since $z_k=0$ for some $k$ implies that $W_p(z)=0.$ 
The next result is a consequence of the fact that the generalized mean is non decreasing (see for instance  \cite{b03} and \cite{bdm25}). 
 
\begin{prop}\label{monotoneNew}Suppose that $p,q\in \Real$, and let  $W_{{p}}$ and $ W_{{q} }$ be two generalized mean utility functions defined on $\Real_+^d$. If $p\geq q$, then for all $\delta\in \Real_+^d$
$$W_{{p}}(\delta)\geq W_{{q}}(\delta).$$
Moreover let $\bar p\in \Real\cup\{-\infty,+\infty\}$ and suppose that $\{p_k\}_{k\in \mathbb N}$ is a monotone sequence of real numbers such that $\lim_{k\longrightarrow \infty}p_k=\bar p$ and $\{W_{{p_k}}\}_{k\in \mathbb N}$ is a sequence of generalized mean  utility functions that is pointwise convergent   to $W_{\bar p}$. Then, $\{W_{{p_k}}\}_{k\in \mathbb N}$ converges uniformly on any compact subset $  K$ of $\Real_+^d$. In addition:
$$\lim_{ k\longrightarrow \infty  }\mathscr D_{W_{{p_k}}}=\mathscr D_{W_{{\bar p} }}$$
 
\end{prop}

In line with the afore mentioned properties, it would also be interesting to consider a function derived from the Kolm-Pollak form arising in social choice theory (see \cite{k76}). Let us consider the map  $W_{{p}}^{\ln}$ defined as:
\begin{equation}
W_{{p}}^{\ln}(z)=\frac{1}{p} \ln\Big(\frac{1}{d}\sum_{ k\in [d]}\exp({p z_k})\Big). 
\end{equation}
It is easy to check that this map satisfies the conditions required in Definition \ref{DefUbased} and in particular:

\begin{equation}
\lim_{p\longrightarrow \infty}W_{{p}}^{\ln}(z)=W_{\infty} (z) 
\end{equation}for all $z$. Moreover, $W_{{p}}^{\ln}$ is translation homothetic in the direction of the unit vector:

\begin{equation}
W_{{p}}^{\ln}(z)(z+\delta 1\!\!1_d)=W_{{p}}^{\ln}(z)+\delta. 
\end{equation}


\subsection{From Utility Based Distance Functions and Generalized Means to   Directional Distance Functions}
\label{sec:powermin}
 In the following we  show that the formalism proposed in   section \ref{UForm}   includes as a special case a large number of technical efficiency measures. One can then retrieve   indirectly retrieve as a special case, for every feasible production vectors, all  the afore mentioned directional distance functions.

 For any nonempty subset $ \mathcal K$  of $[d]$, let us denote \begin{equation}\Real_{ \sss{\mathcal K} }=\Big\{\sum\limits_{k\in   \mathcal K}w_ke_k :w_k\in \Real, k\in   \mathcal K \Big\}.\end{equation}
For all vectors $z$ of $\Real^d$, let $z_{\sss{\mathcal  K}}=\sum_{k\in  K}z_ke_k$ denote  the canonical projection of $z$ onto $\Real_{\sss {\mathcal K}}$. For all $g\in \Real_+^{d}\backslash\{0\}$, it follows that if $  \mathcal K=\mathcal G$, then $\Real_{\mathcal G}=\big\{\sum\limits_{k\in  \mathcal G}w_ke_k :w_k\in \Real, g_k>0  \big\}.$ Moreover,  $z_{\mathcal G}=\sum_{g_k>0}z_ke_k$. Clearly, if $g>0$ then $\Real_{\mathcal G}=\Real^{d}$ and $z_{\mathcal G}=z$. Accordingly, let   $\Real_{\sss{\mathcal K},+}$ denotes the set of the nonnegative canonical projections onto $\Real_{\sss {\mathcal K}}$. 

 In the following, we slightly extend the generalized  $p$-mean utility functions introduced in section \ref{UForm} in the case where $g\not=0.$

\begin{enumerate}  \item[(i)] $W_{{p},g}(\delta)=\frac{1}{{| {\mathcal G}|}^{\frac{1}{p}}}\stackrel{\phi_p}{\sum\limits_{k\in
\mathcal G}}\,\frac{\delta_k}{g_k} $ if $p\in \Real\backslash\{0\}$;\\

\item [(ii)]   $W_{{0},g}(\delta)=\prod\limits_{k\in \mathcal G} \big(\frac{\delta_k}{g_k}\big)^{\frac{1}{|\sss{\mathcal G}|}} $ if $p=0$;\\
 
\item [(iii)]     $W_{-\infty,g}(\delta)=\min\limits_{k\in \mathcal G}  \frac{\delta_k}{g_k} $ if $p=-\infty$; \\
  
 \item [(iv)]    $W_{\infty,g}(\delta)=\max\limits_{k\in \mathcal G} \frac{\delta_k}{g_k} $ if $p=\infty$.\\
  \end{enumerate} 
 Clearly for all $p\in [-\infty,+\infty]$, $W_{{p},g}$ satisfies over $\Real^{d}$ the conditions of Definition \ref{DefUbased}. It is shown in the following that this formalism captures the notion of direction involved in many existing efficiency measures.

This new  { generalized}   mean directional distance function can be defined as follows:

\begin{defn}\label{powermin}  Under $T1$ to $T4$, for all  $p\in \Real\cup\{-\infty,+\infty\}$ and all $g\in \Real_+^d\backslash\{0\}$, the generalized mean directional distance function is the map $\mathscr D_{W_{{p},g}}:\Real_{\mathcal G,+}\longrightarrow \Real_+\cup \{-\infty,+\infty\}$ defined as:$$\mathscr D_{W_{{p},g}}(z)= \sup\limits_{u }\Big\{ W_{{p},g}(u-z) : u\in T_z \Big\}.$$

\end{defn}

By construction,  following Definition \ref{DefUbased}, $\mathscr D_{W_{{p},g}}$ admits a $W_{{p},g}$ utility based representation for any $p$. {In  Definition \ref{powermin}, we consider the cases where $p$ takes either positive or negative values}. The number  $\frac{1}{{|\mathcal G|}}$   operates as a weighting scheme taking into account the dimensions of the netput space. 

 The next statement shows that the generalized  mean directional distance function encompasses as a special case many distance functions  including the directional F\"are-Lovell distance function. In addition, it relates the directional distance function to the map $\delta \mapsto \min_{k\in \mathcal G}\delta_k$.

\begin{prop}\label{maxfar}
$T1$ to $T4$, for all $p\in \mathbb R\backslash\{0\}$, the  {generalized} mean directional distance function   $ \mathscr D_{W_{{p},g}}: \Real^d  \longrightarrow \Real_+\cup \{-\infty,+\infty\}$ satisfies the following relations for all $z\in T$: 

\noindent $(a)$ If $p\in \Real\backslash \{0\}$ then:
\begin{align*} 
 \mathscr D_{W_{{p},g}}(z)= \sup\limits_{\delta\in \Real_+^d}\Big\{ \frac{1}{{| \sss { \mathcal G}|}^{\frac{1}{p}}}
\stackrel{\phi_p}{\sum_{k\in \mathcal G}}\delta_k : z+\delta \odot g\in T \Big\}.
\end{align*}
\noindent $(b)$ If $p=0$ then $\mathscr D_{W_{{0},g}}(z)=    \sup \limits_{\delta
\in \Real_+^{d}}   \Big\{ \prod \limits_{k\in
\mathcal G} \delta_k^{\frac{1}{|\scriptscriptstyle G|}} :   z+\delta \odot g \in T  \Big\}.$

\noindent $(c)$ If $p=-\infty$ then $\mathscr D_{W_{-\infty,g}}(z)=\mathrm{D}(z;g).$

\noindent $(d)$ If $p=\infty $ then  $\mathscr D_{W_{\infty,g}}(z)  =\mathrm{AD}(z;g) .$

\end{prop}

The next result is an immediate corollary:

\begin{cor}Under $T1$ to $T4$, for all  $p\in \Real\cup\{-\infty,+\infty\}$ and all $g\in \Real_+^d\backslash\{0\}$,
$$\mathscr D_{W_{{p},g}}(z)=   \sup\limits_{\delta\in \Real_+^d}\Big\{W_{{p},g}(\delta) : z+\delta \odot g\in T \Big\}.
$$
\end{cor}

 Proposition \ref{maxfar} shows that:\\

 $(i)$ One retrieve the directional distance function when $p=-\infty.$ In  particular: \begin{equation}\mathscr D_{W_{-\infty,g}}(z)= \sup\limits_{ u\in T_{z}} \min\limits_{k\in \mathcal G}   \big\{\frac{u_k-z_k}{g_k}\big\}=\mathrm{D}(z;g).\end{equation}

$(ii)$  If $p>0$, then we have:
 \begin{align}
\mathscr D_{W_{{p},g}}(z) = \sup\limits_{\delta\in \Real_+^d}\Big\{      \big( \frac{1}{|\sss{\mathcal G}|}\sum\limits_{k\in
\mathcal G} {\delta_k}^p    \big)^{1/p} \!\! :  z+\delta \odot g \in T\Big\} . \end{align}

$(iii)$ From   \eqref{DFL}, we retrieve the directional F\"are-Lovell measure with:
\begin{align}
 \mathscr D_{W_{{1},g}}(z)= \sup\limits_{\delta\in \Real_+^d}\Big\{       \frac{1}{|\sss{\mathcal G}|}\sum\limits_{k\in
\mathcal G} {\delta_k}     \!\! :  z+\delta \odot g \in T\Big\}= \mathrm{D}_{\mathrm{FL}}(z;g)=\mathscr D_{W_{{1}}}(z) . \end{align}

$(iv)$ The case $p=0$ yields a   multiplicative
directional F\"are-Lovell distance function extending a multiplicative analogue F\"are-Lovell measure   proposed in \cite{rb98}. 

$(v)$ Finally the asymmetric directional distance function can be reformulated as:
\begin{equation}
\mathscr D_{W_{\infty,g}}(z)=\sup\limits_{ u\in T_{z}} \max\limits_{k\in \mathcal G}   \big\{\frac{u_k-z_k}{g_k}\big\}=  \mathrm{AD}(z;g).
\end{equation}

{It is now shown that the generalized mean directional distance function is related to an optimal reference netput vector. These elementary properties are important for the results established hereafter in the paper.}

\begin{lem} \label{compact}
Under $T1$ to $T4$, for all $z\in T$, $g\in \Real_+^d\backslash\{0\}$ and   all $p\in \Real\cup\{-\infty,+\infty\} $, there is some 
 $  \delta_{{p}}^\star \in \Real_{\mathcal G,+}$ such
that $$\mathscr D_{W_{{p},g}}(z) =W_{{p},g}( \delta_{{p}}^\star) $$ with  $ z+\delta_{{p}}^\star \odot g \in  \mathcal F(T).$ Moreover, if $p\in \Real_{++}$, then $ z+\delta_{{p}}^\star \odot g \in \mathcal E(T).$
\end{lem}

Note that if $p<0$ then  $ z+\delta_{{p}}^\star \odot g$ may not be in $\mathcal E(T)$. This comes from the fact that the map $\delta\mapsto \stackrel{\phi_p}{\sum\limits_{k\in [d]}}\delta_k$ is not strictly increasing at point $0$. $0$ is an absorbing element of the generalized sum.

\subsection{Axiomatic Properties}
\label{subsec:prop}

The generalized mean directional distance function involves a preassigned direction \( g \). In what follows, we characterize its axiomatic properties.  This is done in the next proposition by introducing a suitable refinement of the concept of an efficient subset.

For all  $k\in [d]$ let us   define the $k$-input weak efficient subset as:
\begin{equation}
\mathcal F_k(T)=\Big \{ z\in T: z'\geq z \text{ and }z_k>z'_k\Longrightarrow z'\notin T\Big \}.
\end{equation}
This definition relaxes the standard definitions of weak efficiency by focusing on a single specific commodity $k$. Briec, Cavaignac and Kerstens \cite{bck11} propose a similar idea  with the notion of {\it weak efficient subset in the direction of $g$}. The proof of the next statements is similar and thus is omitted.

The efficient subset of $T$ satisfies the above criteria for all   $k\in [d] $ and the weak efficient subset statisfies it at least for some $k$. Therefore, they can be expressed as:
\begin{equation}
\mathcal E(T)= \bigcap_{k\in [d]}\mathcal F_k(T)\quad \text{ and } \quad \mathcal F(T)=\bigcup_{k\in [d]} \mathcal F_{ k}(T).
\end{equation}
   
For all subsets $\mathcal K$ of $[d]$   let us denote:
\begin{equation}
\mathcal E_{\mathcal K}(T)= \bigcap_{k\in \mathcal K}\mathcal F_{k}(T) .
\end{equation}
$\mathcal E_{\mathcal K}(T)$ is termed the {\bf $\mathcal K$-efficient subset}. This definition   weaken    the usual notion of efficient subset by focusing of some specific set of netputs. This subset is  characterized from the generalized F\"are-Lovell directional function.  

  Paralleling these directional definitions of the weak and strong efficient subsets in the direction of $g$, we say that $f$ is {\bf strongly decreasing in the direction of $g$} if $z_{\mathcal G}\not=z'_{\mathcal G}$ and $z'\geq z$ imply that 
$f(z')<f(z)$. $f$ is {\bf  weakly decreasing  in the direction of $g$}, if $f$ is non increasing and if  $z'_{\mathcal G}>  z_{\mathcal G}$ implies that 
$f(z')<f(z)$.

\begin{prop}\label{AXPROP}
\sloppypar{Under $T1$ to $T4$,  {for all $p\in \Real$} and all $g\in \Real_+^d\backslash\{0\}$, we have:}\\
$(a)$ If $p>0$, then $\mathscr D_{W_{{p},g}}(z)=0$ if and only if $z\in \mathcal E_{\mathcal G}( T)$.\\
$(a')$ If $p\leq 0$, then $\mathscr D_{W_{{p},g}}(z)=0$ if and only if $z\in \mathcal F_{\mathcal G}( T)$.\\
$(b)$   For all  $p\in \Real$
we have  $ \mathscr D_{W_{{p}}}(z)  $ is non increasing. \\
\noindent $(c)$  For all $p>0$, $\mathscr D_{W_{{p},g}}(z)$ is strongly decreasing in the direction of $g$.\\
  \noindent $(c')$  For all $p\leq 0$, $\mathscr D_{W_{{p},g}}(z)$ is weakly decreasing in the direction of $g$.\\
$(d)$ If $L$ is a $d\times d$  positive diagonal matrice  then we have the equality
$\mathscr D_{W_{{p},g}}(z)= \mathscr D_{W_{{p},Lg}}(Lz)$.\\
\end{prop}
{Notice that many of these axioms are for instance discussed \cite{fl78}, \cite{r85,r88} of in \cite{chk99}.} Propositions \ref{AXPROP}.$(a)$ and \ref{AXPROP}.$(c)$ show that if $ p > 0 $, the generalized mean directional distance function characterizes the strong efficient subset and is strongly decreasing in the direction of \( g \). Similarly, Propositions \ref{AXPROP}.$(a')$ and \ref{AXPROP}.$(c')$ show that if $ p \leq 0 $, the generalized mean directional distance function characterizes the weak efficient subset and is weakly decreasing in the direction of $ g $. From Proposition \ref{AXPROP}.$(b)$, it follows that the generalized mean directional distance function is nonincreasing in all cases. Proposition \ref{AXPROP}.$(d)$ establishes that the generalized mean directional distance function is independent of the units of measurement.{Note that the proposed measure does not satisfy any analogue of the restricted scale invariance generally used in bargainig theory(see for instance \cite{ht10}) because a translation of the production set is not compatible wit the axiom T1.}

Proposition \ref{AXPROP} indicates that the directional F\"are-Lovell distance function satisfies the axiomatic properties inherited from the standard case ($ p = 1 $). However, it is well known that the directional distance function (\( p = -\infty \)) does not characterize the strong efficient subset. This limitation arises because the map $ \delta \mapsto \min\limits_{k \in [d]} \delta_k $ is not strictly increasing. A similar issue occurs with the asymmetric directional distance function derived from the map $ \delta \mapsto \max\limits_{k \in [d]} \delta_k $, which is also not strictly increasing. As a result, maximizing these functions does not enforce the necessary adjustments among netput combinations to reach the efficient subset.

The generalized mean directional distance function extends the properties satisfied by the input directional F\"are-Lovell efficiency measure defined in \cite{bck11}. It characterizes the partial efficient subset for a given direction \( g \). In line with Lemma \ref{compact}.(c), it fully characterizes the efficient subset only when \( p > 0 \) and \( g > 0 \). All these measures clearly belong to the class of slack-based efficiency measures considered in \cite{fw09}.

\subsection{  Directional, Asymmetric and Multiplicative Distance Functions as a Special Limiting Case}
\label{subsec:Limit}

{The directional distance function, the directional F\"are-Lovel distance function, the directional  multiplicative F\"are-Lovell, and the asymmetric directional distance functions can be viewed as a limiting case of the generalized mean directional  distance function. Notice, however, that we only consider the case of feasible netput vectors.  }

\begin{prop}\label{limfunc} Under $T1$ to $T4$,  for all $z\in T$    and all $g\in \Real_+^d\backslash\{0\}$, we have:

\noindent $(a)\quad $ $\lim\limits_{p\longrightarrow -\infty}\mathscr D_{W_{{p},g}}(z)= \mathrm{D}(z; g)$.

\noindent $(b)\quad $ $\lim\limits_{p\longrightarrow 0^-}\mathscr D_{W_{{p},g}}(z)=\mathscr D_{W_{{0}}}(z)$.

\noindent $(c)\quad $ $\lim\limits_{p\longrightarrow +\infty}\mathscr D_{W_{{p},g}}(z)=\mathrm{AD}(z; g)$.
\end{prop}

The radial efficiency measure proposed by Debreu \cite{d51} and
Farrell \cite{f57} is defined as
$\mathrm{E}_{\mathrm{DF}}:\Real^n_-\times\Real_+^m \longrightarrow
\Real_+\cup\{-\infty,\infty\}$ is defined for all $z:=(x,y)\in T$ as follows:
\begin{align}\label{DF}
\mathrm{E}_{\mathrm{DF}}(z) = \inf \limits_{\lambda \in \Real_+}\left\{\lambda: 
(\lambda  x, y)\in T\right\} .
\end{align}
If the condition is not vacuous, that is there is a $\lambda$ with $(\lambda x,y\in T)$ then the sup will be achieved; otherwise it is defined to be $+\infty$. This radial efficiency measure indicates the maximal equiproportionate
reduction in all inputs which still allows production of the given output vector
on the isoquant of the input set.

Chambers, Chung and F\"are \cite{ccf96} showed that $\mathrm{D}(x,y; x,0)=\mathrm{E}_{\mathrm{DF}}(x,y)$.  It follows from Proposition \ref{limfunc} that
\begin{equation}
\lim_{p\longrightarrow -\infty}\mathscr D_{W_{{p},(x,0)}}(x,y)=1-\mathrm{E}_{\mathrm{DF}}(x,y).
\end{equation}

Figures 6 and   7 depict the case where $p<0$. When $p\longrightarrow -\infty$ the map $u\mapsto W_{{p},g}(u-z)=\frac{1}{{|\mathcal G|}^{\frac{1}{p}}}\stackrel{\phi_p}{\sum\limits_{k\in
\mathcal G}}\frac{u_k-z_k}{z_k} $ tends to some kind of netput oriented Leontief function defined  \begin{equation}u\mapsto W_{-\infty,g}(u-z)=\min_{k\in \mathcal G}  \big\{\frac{u_k-z_k}{g_k}\big\} .\end{equation}
\bigskip \begin{center}{\scriptsize

\unitlength 0.4mm 
\linethickness{0.3pt}
\ifx\plotpoint\undefined\newsavebox{\plotpoint}\fi 
 

\begin{picture}(298,164.75)(0,0)
\put(130,35.75){\vector(-1,0){121.75}}
\put(129.75,36.25){\vector(0,1){110.5}}
\put(51.5,84.75){\circle*{1.414}}
\put(51.5,84.75){\line(0,1){60.5}}
\put(51.5,84.75){\line(1,0){80.5}}
\put(145.25,55.5){\vector(3,4){.07}}\multiput(129.75,36)(.0336956522,.0423913043){460}{\line(0,1){.0423913043}}
\qbezier(84.75,113.5)(66.875,128)(59.5,163.5)
\qbezier(126.5,104.25)(92.75,107.625)(85,113.5)
\put(96.5,65.25){\makebox(0,0)[]{$T$}}
\put(76.5,95.25){\makebox(0,0)[cc]{$T_{z}$}}
\put(134.75,31.25){\makebox(0,0)[]{$0$}}
\put(131.5,157){\makebox(0,0)[]{$y$}}
\put(47.75,75.25){\makebox(0,0)[cc]{$z$}}
\put(99.5,123.5){\makebox(0,0)[cc]{$z^{\star}_p$}}
\put(83.75,114.5){\circle*{1.581}}
\put(50,164.75){\makebox(0,0)[]{$p<0$}}
\put(47.25,1.25){\makebox(0,0)[]{{\bf Figure 7:} The case $p\in ]-\infty, 0[$}}
\put(152.25,62.5){\makebox(0,0)[cc]{$g$}}
\put(156,35.25){\vector(-1,0){.07}}\multiput(280.5,35)(-15.5625,.03125){8}{\line(-1,0){15.5625}}
\put(280.5,35.25){\vector(0,1){108.5}}
\put(298,54){\vector(1,1){.07}}\multiput(280.75,35.5)(.0336914063,.0361328125){512}{\line(0,1){.0361328125}}
\put(184.5,83.25){\circle*{1}}
\put(219,120.75){\vector(1,1){.07}}\multiput(184.75,84)(.03371062992,.03617125984){1016}{\line(0,1){.03617125984}}
\put(219,121){\line(1,0){50.5}}
\put(219,121){\line(0,1){43.25}}
\put(238.5,64.25){\makebox(0,0)[]{$T$}}
\put(218.5,94.25){\makebox(0,0)[cc]{$T_z$}}
\put(284.75,32){\makebox(0,0)[]{$0$}}
\put(280.5,151.5){\makebox(0,0)[]{$y$}}
\put(179.75,77){\makebox(0,0)[cc]{$z$}}
\put(200.5,120.25){\makebox(0,0)[cc]{$ z^\star_{-\infty} $}}
\put(218.75,120.75){\circle*{1.581}}
\put(184.5,83.25){\line(0,1){60.5}}
\put(184.5,83.25){\line(1,0){80.5}}
\put(199,158.75){\makebox(0,0)[]{$p=-\infty$}}
\put(209,0){\makebox(0,0)[]{{\bf Figure 8:} The directional case $p=-\infty$}}
\put(298,61){\makebox(0,0)[cc]{$g$}}
\put(0,35.5){\makebox(0,0)[cc]{$x$}}
\put(149,35){\makebox(0,0)[cc]{$x$}}
\put(129.75,43.25){\line(0,-1){34.75}}
\put(280.5,42.5){\line(0,-1){34.75}}
\qbezier(36,137.5)(116,113.75)(130,36)
\qbezier(4.75,157.25)(17,143.5)(35.25,137.75)
\qbezier(280.5,35)(259.125,120.5)(179.25,132)
\qbezier(179.25,132)(159.125,135)(142.5,154)
\end{picture}

}

\end{center}

\begin{center}{\scriptsize

\unitlength 0.4mm 
\linethickness{0.3pt}
\ifx\plotpoint\undefined\newsavebox{\plotpoint}\fi 
 

\begin{picture}(289.25,150)(0,0)
\put(126,27.75){\vector(-1,0){121.75}}
\put(125.75,28.25){\vector(0,1){110.5}}
\put(145.75,27.5){\vector(-1,0){.07}}\multiput(270.25,27.25)(-15.5625,.03125){8}{\line(-1,0){15.5625}}
\put(270.25,27.5){\vector(0,1){108.5}}
\put(47.5,76.75){\circle*{1.414}}
\put(141.25,47.5){\vector(3,4){.07}}\multiput(125.75,28)(.0336956522,.0423913043){460}{\line(0,1){.0423913043}}
\put(285.25,46.25){\vector(3,4){.07}}\multiput(269.75,26.75)(.0336956522,.0423913043){460}{\line(0,1){.0423913043}}
\put(174.25,75.5){\circle*{1}}
\put(0,27.75){\makebox(0,0)[]{$x$}}
\put(272.75,22.75){\makebox(0,0)[]{$0$}}
\put(130.75,23.25){\makebox(0,0)[]{$0$}}
\put(127.5,149){\makebox(0,0)[]{$y$}}
\put(270.25,143.75){\makebox(0,0)[]{$y$}}
\put(43.75,67.25){\makebox(0,0)[cc]{$z$}}
\put(169.5,69.25){\makebox(0,0)[cc]{$z$}}
\put(43.75,1){\makebox(0,0)[]{{\bf Figure 9:} The case $p\in ]  0,+\infty[$}}
\put(229.25,0){\makebox(0,0)[]
{{\bf Figure 10:} The asymmetric  directional case $p=+\infty$}}
\put(91,60){\makebox(0,0)[]{$T$}}
\put(115.5,85.5){\makebox(0,0)[cc]{$ z^{\star}_{p}$}}
\put(259.25,85.5){\makebox(0,0)[cc]{$z^{\star}_{\infty}$}}
\put(47.75,77){\line(0,1){59}}
\put(47.5,77){\line(1,0){67.25}}
\qbezier(107,62.25)(101.875,138.625)(20.25,143.5)
\put(82,84.75){\makebox(0,0)[cc]{$T_z$}}
\put(174,76.25){\line(0,1){64.5}}
\put(174.25,76){\line(1,0){84}}
\put(248,69.5){\line(0,1){79.75}}
\put(248,149.25){\line(-1,0){79.5}}
\put(220.75,84.75){\makebox(0,0)[cc]{$T_z$}}
\put(230.5,48.25){\makebox(0,0)[]{$T$}}
\put(105,76.75){\circle*{1.9}}
\put(247.75,76){\circle*{1.5}}
\put(145.25,50.75){\makebox(0,0)[cc]{$g$}}
\put(289.25,49.5){\makebox(0,0)[cc]{$g$}}
\put(202.25,140.5){\makebox(0,0)[cc]{$p=+\infty$}}
\put(125.75,27.75){\line(0,-1){21}}
\put(270,27.75){\line(0,-1){21}}
\qbezier(126,28)(109,102.375)(41,121.25)
\qbezier(270,25.5)(253,99.875)(185,118.75)
\qbezier(41.25,121.25)(3.25,131.875)(1.25,150)
\qbezier(186,118.5)(148,129.125)(146,147.25)
\end{picture}

}

\end{center}

Figures 8 and Figure 9 depict the case where $p>0$. When $p\longrightarrow +\infty$ the map $u\mapsto W_{{p},g}(u-z)=\frac{1}{{|\mathcal G|}^{\frac{1}{p}}}\stackrel{\phi_p}{\sum\limits_{k\in
\mathcal G}}\,   \frac{u_k-z_k}{g_k} $ tends to the function defined  $u\mapsto W_{\infty,g}(u-z)= \max_{k\in \mathcal G}\big\{\frac{u_k-z_k}{g_k}\big\}$. Notice that for certain $p>0$ there is no evidence that the projection point is an extreme point of $T_{z}$. However, this is the case when $p=\infty$.

\begin{center}{\scriptsize

\unitlength 0.4mm 
\linethickness{0.3pt}
\ifx\plotpoint\undefined\newsavebox{\plotpoint}\fi 

\begin{picture}(302,160)(0,0)
\put(124.75,25.5){\vector(-1,0){121.75}}
\put(272.75,26){\vector(-1,0){121.75}}
\put(124.5,26){\vector(0,1){110.5}}
\put(272.5,26.5){\vector(0,1){110.5}}
\put(140,45.25){\vector(3,4){.07}}\multiput(124.5,25.75)(.0336956522,.0423913043){460}{\line(0,1){.0423913043}}
\put(288,45.75){\vector(3,4){.07}}\multiput(272.5,26.25)(.0336956522,.0423913043){460}{\line(0,1){.0423913043}}
\put(47.25,151){\makebox(0,0)[]{$p=1$}}
\put(191,155.25){\makebox(0,0)[]{$p=1$}}
\put(302,26){\makebox(0,0)[cc]{$x$}}
\put(129.5,21){\makebox(0,0)[]{$0$}}
\put(277.5,21.5){\makebox(0,0)[]{$0$}}
\put(126.25,146.75){\makebox(0,0)[]{$y$}}
\put(274.25,147.25){\makebox(0,0)[]{$y$}}
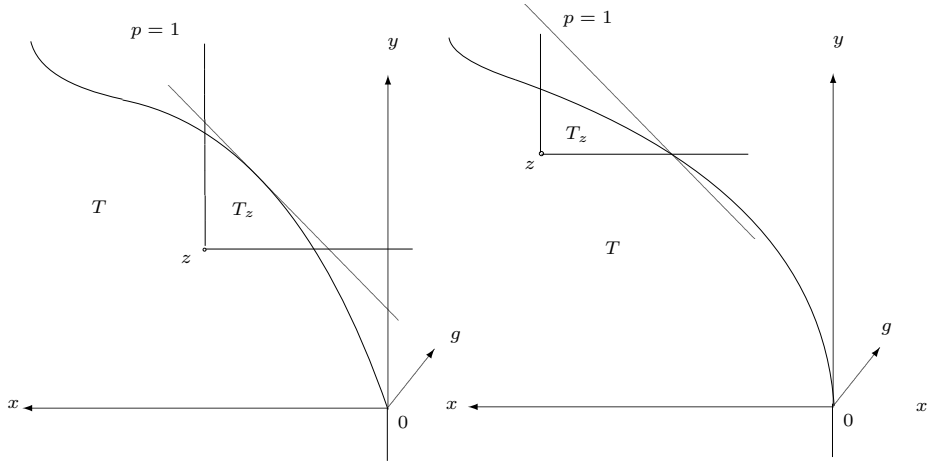
\put(60.75,1.25){\makebox(0,0)[]{{\bf Figure 11:} The case $p=1$}}
\put(200.5,0){\makebox(0,0)[]{{\bf Figure 12:} Kink solution in the  case $p=1$}}
\put(147,49.25){\makebox(0,0)[cc]{$g$}}
\put(175.5,110.25){\circle*{1.414}}
\put(76.5,91.5){\makebox(0,0)[cc]{$T_z$}}
\put(187.5,116.75){\makebox(0,0)[cc]{$T_z$}}
\multiput(63.75,79.5)(-.03125,8.40625){8}{\line(0,1){8.40625}}
\put(64,78.5){\line(1,0){68.75}}
\put(175.5,110){\line(1,0){68.75}}
\multiput(51.5,133)(.03373015873,-.034501763668){2268}{\line(0,-1){.034501763668}}
\multiput(170,160)(.03373015873,-.034501763668){2268}{\line(0,-1){.034501763668}}
\put(63.5,78.25){\circle*{1}}
\put(28.5,92.5){\makebox(0,0)[]{$T$}}
\put(57.5,75.25){\makebox(0,0)[cc]{$z$}}
\put(171.75,106.25){\makebox(0,0)[cc]{$z$}}
\put(175.25,110.5){\line(0,1){39.5}}
\put(199.25,78.75){\makebox(0,0)[]{$T$}}
\put(145.75,26.5){\makebox(0,0)[cc]{$x$}}
\put(0,27){\makebox(0,0)[cc]{$x$}}
\put(290.25,52){\makebox(0,0)[cc]{$g$}}
\put(124.25,25.75){\line(0,-1){17.75}}
\put(272.25,27){\line(0,-1){17.75}}
\qbezier(124.5,25)(92.625,117)(36.25,128)
\qbezier(36.25,128.25)(9,134.125)(5.75,147.5)
\qbezier(272.75,26.25)(267.75,99)(166.75,134.75)
\qbezier(166.75,134.75)(145.75,141.75)(144.75,148.75)
\end{picture}

}
\end{center}
Figures 10 and   11 depict the case $p=1$ and show that the projection   of $z$ onto the efficient frontier may not be a kink point of $T_z$.\\

 Thus, the four types of directional distance functions  discussed in Section
\ref{sec:basic} are clearly limiting cases of the new  {
generalized directional  mean directions distance function } in Definition \ref{powermin}. Note that in the limit cases $p\longrightarrow \infty $, $p\longrightarrow -\infty $ and $p\longrightarrow 0 $, the limit distance functions do not characterize the efficient subset $\mathcal E(T)$.

{Proposition~\ref{limfunc} shows that the generalized mean distance function admits both the asymmetric directional distance function and the directional distance function as limiting cases. The asymmetric case is illustrated in Figure~10. This construction is closely related to the notion of an ideal point introduced in \cite{z74} and was also used in \cite{ht10} to select benchmark efficient points in a production setting.  Figure~8 illustrates the connection between the directional distance function and the maximin approach, which also arises in bargaining theory. In particular, it corresponds to the Kalai--Smorodinsky solution \cite{ks75}, where the outcome is determined by the best attainable payoff of each player. This solution satisfies the axioms of Pareto efficiency, symmetry, maximin fairness, and invariance with respect to linear transformations.  In our framework, however, the ideal point
$m(z; g) = \sum_{k \in \mathcal G} \mathrm{D}(z; g_k e_k)\, e_k$ may differ from the direction vector $g$, since it depends on the underlying technology, whereas $g$ is fixed \emph{a priori}. Finally, the case $p = 0$, obtained as the limit when $p \to 0_-$ and depicted in Figure~7, corresponds to the Nash solution \cite{n50}, which satisfies the axiom of independence of irrelevant alternatives.} 

Paralleling \cite{bck20},     all the generalized mean directional distance functions are ordered as follows. 

 {\begin{prop}\label{Order}Under $T1$ to $T4$, the generalized { mean } directional distance function
can be ordered as follows for all $z\in T$ and   $g\in \Real_+^d\backslash\{0\}$. For all  $p,q\in\Real$, if $p\geq q$ then:
\begin{align*}
\mathrm{D}(z;g)=\mathscr D_{W_{-\infty,g}}(z) \leq \mathscr D_{W_{{q},g}}(z) &\leq \mathscr D_{W_{{p},g}}(z)\leq \mathscr D_{W_{\infty,g}}(z)=\mathrm{AD}(z;g).
\end{align*}
\end{prop}}

Note that the direction of these inequalities is reversed in the approach proposed by \cite{bck20}, where the authors demonstrate that the efficiency measures originally introduced by Debreu \cite{d51} and Farrell \cite{f57} (as $p \to \infty$), F\"are and Lovell \cite{fl78} (for $p = 1$), the asymmetric F\"are measure (as $p \to -\infty$), and the multiplicative F\"are-Lovell measure (as $p \to 0$), are all input-oriented efficiency measures. These can be viewed as particular cases of a newly introduced class of extended F\"are-Lovell input efficiency measures. Specifically, they define a measure $\mathrm{E}_{{p}}: \mathbb{R}^m_- \setminus \{0\} \times \mathbb{R}_+^n \to [0,1] \cup \{\infty\}$, which for any $z := (x, y) \in T$ is given by:

\begin{align}
\mathrm{E}_{{p}}(x,y) = \inf\limits_{\beta \in [0,1]^n} \left\{ \frac{1}{|\sss {\mathcal I(x)}|^{1/p}} \sum_{i \in \sss {\mathcal I(x)}}^{\phi_p} \beta_i : (\beta \odot x, y) \in T \right\}.
\end{align}

If $z \notin T$, the value of the measure is defined to be $+\infty$. This measure exhibits a multiplicative structure and is input-oriented.

It is further shown that, for all $q \geq p > 0$,
\begin{equation}
\mathrm{E}_{\mathrm{DF}}(z) \geq \mathrm{E}_{{p}}(z) \geq \mathrm{E}_{ q }(z) \geq \mathrm{E}_{\mathrm{FL}}(z) \geq \mathrm{AF}(z),
\end{equation}
where $AF$ denotes the asymmetric F\"are measure. Further research may be pursued from an axiomatic perspective, particularly by analyzing the continuity properties of the various proposed efficiency measures (see \cite{rs18}).


\section{Profit Function Prices and Duality for Convex and Non-Convex Technologies}
\label{sec:Dual}
This section establishes  several   results connecting the generalised mean directional distance function and the profit function. First, a duality theorem is established for the maximisation of a  distance over a compact subset in a normed space. In \cite{bck20} the Nirenberg's theorem (that deals with the minimisation of the distance to a convex set) is used establish a duality result relating the input generalized F\"are-Lovell measure $\mathrm{E}_{{p}}$ and the cost function. An analogue maximization theorem is  proposed to relate the generalized F\"are-Lovell distance function and the profit function when $p\geq 1$ (see \cite{l68} for a proof of the Nirenberg's Theorem). Notice, however, that convexity is not required. In the case where $p<1$ the notion of norm is no longer suitable in our context. However,  a   useful duality result   is derived for quasi-concave utility-based distance functions.  A suitable duality result is then obtained using standard rules of Lagrangian duality. In the latter case, the formal rules involved in our duality results parallel the formal relationships existing between $\ell_p$ and $\ell_q$ norms with the condition $\frac{1}{p}+\frac{1}{q}=1$.
\subsection{Maximum Norm Distance Functions and $\phi_p$-generalized Sums for $p\geq 1$.}

In the following we consider a real normed vector space $E$ equipped with a norm $\|\cdot\|$. The dual space of $E$ is the set $E^\star$ of all the continuous linear forms defined on $E$ (that is the topological dual of $E$). The dual norm of a linear form
$\varphi_y: x\mapsto \langle y,x\rangle $ is defined as
\begin{equation}
\|\varphi_y\| =\sup\Big \{\frac{|\langle y,x\rangle|}{\|x\|}: x\not=0\Big \}=\|y\|^\star.
\end{equation}
The following result does not require the convexity of $C$. This is due to the fact that the largest ball centered at $ z$ which contains $C$, admits at its maximum point a supporting hyperplane which is also that of $C$. The concept of farest distance we intially considered in \cite{l75}. For the proof of this result see for instance \cite{s07} (see also \cite{nnt23} for some related results).

\begin{prop}\label{maxnorm}Let $C$ be a compact subset of  a real normed vector pace $E$. Let $h_C: E^\star\longrightarrow \Real\cup \{-\infty,\infty\}$ be the functional support of $C$.  Then for all $x\in C$,
$$\max \{\|x-z\|: x\in C\}=\sup_{y\in E^\star} \{h_C(y)-\langle y,z\rangle: \|y\|^\star=1\}. $$

\end{prop}

Suppose that $E$ is the Euclidean vector space $\Real^d$ endowed with the $\ell_p$-norm defined by $x\mapsto \big(\sum_{k\in [d]}|x_k|^p\big)^{\frac{1}{p}}$ if $p\in [1,\infty[$ and as  $x\mapsto  \max_{k\in [d]}|x_k| $ if $p=\infty$. Given any vector $y\in \Real^d$ the dual norm of the map
$\varphi_y: x\mapsto \langle y,x\rangle=\sum_{k\in [d]} y_kx_k $ is defined as
\begin{equation}
\|\varphi_y\|_{p} =\sup\Big \{\frac{|\langle y,x\rangle|}{\|x\|_p}: x\not=0\Big \}=\|y\|_q,
\end{equation}
with $\frac{1}{p}+\frac{1}{q}=1$.

\begin{cor}Let $C$ be a compact subset of  $\Real^d$. Let $h_C: \Real^d\longrightarrow \Real\cup \{-\infty,\infty\}$ be the functional support of $C$.  Then for all $z\in C$ and all $p,q\in [1,\infty]$ with $\frac{1}{p}+\frac{1}{q}=1$,
$$\max \{\|x-z\|_p: x\in C\}=\sup_{y\in \Real^d} \{h_C(y)-\langle y,z\rangle: \|y \|_q=1\}. $$

\end{cor}

Let us define as $\Pi_{z}: \Real_+^{d} \longrightarrow \Real \cup \{\infty\}$ the function which yields the maximum
profit for all the netput vectors of $T_z$. Namely,

\begin{equation}\Pi_{z}(w )=\sup\{w.u :u\in T_z\}.\end{equation}

In the following we consider the situation of a $W$-utility-based distance function where $W$  is a  norm $\|\cdot\|$ that is weakly increasing over the nonnegative orthant $\Real_+^d$. We have the following property.

\begin{prop}\label{normprod}Suppose that $T$ satisfies $T1$ to $T4$. For all $z\in T$, if $W=\|\cdot\|$   is weakly increasing over $\Real_+^d$, then:
$$\mathscr D_{\|\cdot\|}(z)=\max \{\|u-z\|: u\in T_z\}=\sup_{w\geq 0} \{\Pi_z(w)-w.z: \|w\|^\star=1\}. $$
For all $p\in [1,\infty]$ if $W=\|\cdot\|_p$ then 
$$\mathscr D_{\|\cdot\|_p}(z)=\max \{\|u-z\|_p: u\in T_z\}=\sup_{w\geq 0} \{\Pi_z(w)-w.z: \|w\|_q=1\} $$
with $\frac{1}{p}+\frac{1}{q}=1.$

\end{prop}
Not that Proposition \ref{normprod} does not assume $T5$. A utility-based distance function can be constructed from any homogeneous and quasi-convex increasing  function. In such a case it is  however  easy to show that it   necessarily boils down to a norm.

\subsection{Quasi-Concave Utility-Based Distance Functions  and $\phi_p$-generalized Sums for $p< 1$}

In this subsection, we establish a  duality result for a large class of a $W$-utility-based distance functions. 
It should  however  be assumed that $W$ is  positively homogeneous of degree 1. Let us consider the indirect utility function $W^\star:\Real_+^d\longrightarrow \Real\cup\{+\infty\}$ defined as:

\begin{equation}
W^\star(w)=\sup_{z\geq 0}\{W(z):w.z=1\}. 
\end{equation}

It follows from the homogeneity of  $W$ that:
\begin{equation}
\alpha W^\star(w)=\sup_{w\geq 0}\{  W(\alpha z):w.z=1\}=\sup_{w\geq 0}\{  W(z'):w.z'=\alpha\}.
\end{equation}

 A subset $C$ of $\Real_+^d$ is comprehensive if for all $z\in C$, $0\leq z'\leq z$ implies that $z'\in C$.

 \begin{lem}\label{strictpos}Let $C$ be a compact convex comprehensive subset of $\Real_+^d$ that contains $0$. Then for every $z\notin C$, there is a positive vector 
 $w_z\in \Real_{++}^d$ such that $w_z.z>h_C(w_z)$. 
 $$C=\bigcap_{w\in \Real_{++}^d}\{z: w.z\leq h_C(w)\}. $$
\end{lem}
Lemma \ref{strictpos} is useful to simplify the technical exposition of the paper and computing the indirect utility-based distance functions involved by the generalized means when $p<1$. Notice that if $w\in \Real_{++}^d$, then $W^\star(w)<+\infty.$

The following duality result holds under the assumption that $W$ is quasi-concave. For a comprehensive treatment of duality results derived under broad and minimal assumptions, see, for example, \cite{ml91}. 

\begin{prop}\label{quasiconcdual}Let  $W:\Real^d\longrightarrow \Real$   such that $(i)$ $W(0)=0$; $(ii)$ $W$ is upper semi-continuous ; $(iii)$ $W$ is strongly increasing; $(iw)$ quasi-concave. Suppose moreover that $W$ is  positively homogeneous of degree 1. Under $T1$ to $T5$, the   $W$-utility-based distance function $\mathscr D_W$ satisfies the following dual property:
$$\mathscr D_W(z)=\inf_{w\geq 0}\{\Pi_z(w)-w.z:  W^\star(w)=1\}. $$

\end{prop}

 The following example illustrates a situation where the dual solution is not reached under the normalization constraint $W^\star(w)=1$. 
 \begin{expl}\label{exudemi}Let us consider the production set $T=\{(x_1,x_2): x_1\leq 0, x_1+x_2\leq 0, x_2\leq 2\}$. Let us consider the points $z=(-3,2)$ and $z^\star=(-2,2)$. 
 $z^\star$ is efficient and $z$ is weakly efficient. Suppose that $W(u)=(u_1^{\frac{1}{2}}+u_2^{\frac{1}{2}})^2$. In such a case  $W^\star(w)=(w_1^{-\frac{1}{3}}+w_2^{-\frac{1}{3}})^3$. $\mathscr D_{W}(-3,2)= (0+1)^2=1$. 
 We have $T_z=\{(u_1,u_2): u_1\in [-3,-2], u_2=2\}$. However there does not exists a normalized supporting hyperplane that achieves the dual program having null components. If either $w=(1,0)$ or  $w=(0,1)$, we have from the absorption property of the generalized sum $W^\star(w)=+\infty$. However, let us consider for all   $n\in \mathbb N$ the vector $w^{(n)}=(1, {n})$.  We have $W^\star(1, {n})=(1+\frac{1}{n^3})^{3}$, and $\Pi(w^{(n)})-w^{(n)}.z=-2+3=1$. Finally, $$\mathscr D_W(-3,2)=1=\lim_{n\longrightarrow \infty}\big(1+\frac{1}{n^3}\big)^{3}=\lim_{n\longrightarrow \infty}\big (\Pi(w^{(n)}\big )-w^{(n)}.z)W^\star(w_n). $$ \end{expl}

In the next example, there is a normalized price vector that is the solution of the dual problem.  \begin{expl}Let us consider the production set $T$ defined in Example \ref{exudemi} with the point $z=(-3,2)$ that is weakly efficient. Suppose that $W(u)=(u_1^{-\frac{1}{2}}+u_2^{-\frac{1}{2}})^{-2}$. In such a case  $W^\star(w)=(w_1^{ \frac{1}{3}}+w_2^{ \frac{1}{3}})^{-3}$. From the absorption property of the generalized sum, we have $\mathscr D_{W}(-3,2)= 0$. 
If $w=(0,1)$, $\Pi_z(w)=2$ and $W^\star(0, 1)=(0+1)^{-3}=1$. Moreover $\Pi(w)-w.z= 2-2=0$ and $\mathscr D_W(-3,2)=0= (\Pi(w  )-w .z)W^\star(w). $ \end{expl}

\begin{rem}In the case where $p=1$, the map $w\mapsto \sum_{k\in [d]}w_i$ is concave and convex over $\Real_+^d$.  Therefore Proposition \ref{normprod} and \ref{quasiconcdual} apply. Moreover, this map boils down to the $\ell_1$-norm over $\Real_+^d$. From Proposition \ref{normprod}
\begin{equation}
\mathscr  D_{\|\cdot\|_1}(z)=\sup_{w\geq 0}\Big\{\Pi(w)-w.z: \max_{k\in [d]}|w_k|=1\Big\}.
\end{equation}
A solution of the dual problem is a price that defines a supporting hyperplane of the $\ell_1$-ball $B_1(z,\alpha]$ with $\alpha=\mathscr  D_{\|\cdot\|_1}(z)$. Due to the geometric structure of the $\ell_1$-ball we have $w^\star=1\!\!1_d$. 
However, if $W=\|\cdot\|_1$, then for every price vector $w>0$, we have $W^\star(w)= \max_{k\in [d]}{w_k}^{-1}$. Therefore
\begin{equation}
\mathscr D_{W}(z)=\inf_{w\geq 0}\Big\{\Pi(w)-w.z: \max_{k\in [d]}{w_k}^{-1}=1\Big\}=\inf_{w\geq  0}\Big\{\Pi(w)-w.z: \min_{k\in [d]}{w_k}=1\Big\}.
\end{equation}
It follows that the normalization condition $\min_{k\in [d]}{w_k}=1$ ensures that $ \Pi(w)-w.z $ is minimum for $w^\star=1\!\!1_d$ and we find the same result $\mathscr D_{W}(z)=\mathscr D_{\|\cdot\|_1}(z)$ with two different dual programs. 

\end{rem}

In the following we consider the case where $p<1$. In such a case the map $v\mapsto \stackrel{\phi_p}{\sum_{k\in [d]}}v_k$ is not immediately related to the standard notion of norm. The next result is useful to connect the generalized mean directional distance function to the profit function. Along this line, a duality result is derived from Proposition \ref{quasiconcdual}.

From Lemma \ref{maxlem} (see appendix for the statement and the proof and \cite{mwg95} for similar properties), we have the following situations. If $W(v)=\stackrel{\phi_p} {\sum\limits_{k\in [d]}}a_kv_k$ for all $v\in \Real_+^d$ and $a,w\in \Real_{++}^d$ then, for all $p<1$, with $p\not=0$, we have:
\begin{align}
W^\star(w)=\Big[\stackrel{\phi_q}{\sum_{k\in [d]}}a_k^{-1} {w_k}\Big]^{-1},
\end{align}
where $\frac{1}{p}+\frac{1}{q}=1$.{In particular, setting $s = \phi_p(a)$ yields
$W(v) = \phi_p^{-1}\!\left( \sum_{k \in [d]} s_k \phi_p(v_k) \right)
= \stackrel{\phi_p}{\sum\limits_{k \in [d]}} \phi_p^{-1}(s_k)\, v_k,$  and
\begin{equation}\label{trans}
W^\star(w)
= \Big[ \stackrel{\phi_q}{\sum_{k \in [d]}} s_k^{-\frac{1}{p}} w_k \Big]^{-1}
= \Big[ \stackrel{\phi_q}{\sum_{k \in [d]}} s_k^{\frac{1-q}{q}} w_k \Big]^{-1}. 
\end{equation}}

If $W(v) = \prod_{k \in [d]} v_k^{t_k}$ for all $v \in \mathbb{R}^d$, with $\sum_{k \in [d]} t_k = 1$, then $W$ is positively homogeneous and
\begin{equation}\label{transcobb}
W^\star(w) = \prod_{k \in [d]} \big( t_k\, w_k^{-1} \big)^{t_k}.
\end{equation}

{Note that when $p \in ]0,1[$, the function $W_p$ coincides with a quasi-norm. In this case, the associated unit ball is nonconvex and the triangle inequality fails to hold. Such quasi-norms are sometimes employed in the context of cumulative prospect theory \cite{tk92}. }
The next result is then an immediate consequence of Proposition \ref{quasiconcdual} and Lemma \ref{maxlem} (see appendix).

\begin{prop}\label{dualneg}Suppose that the production set satisfies $T1-T5$ and let   $a,w\in \Real_{++}^d$.\\
$(a)$ Suppose that
 $W(v)=\stackrel{\phi_p} {\sum\limits_{k\in [d]}}a_kv_k$ for all $v\in \Real_+^d$. Then, for all $p<1$ with $p\not=0$,  we have:
\begin{align*}
\mathscr D_W(z) =\inf \Big \{\Pi_z(w)-w.z:\stackrel{\phi_q}{\sum_{k\in [d]}}a_k^{-1} {w_k}=1, w\in \Real_{+}^d\Big \},
\end{align*}
where $\frac{1}{p}+\frac{1}{q}=1$. \\
\noindent $(b)$ If $t\in \Real_{++}^d$, $\sum_{k\in [d]}t_k=1$, and $W(v )={\prod\limits_{k\in [d]}}(t_kv_k) ^{t_k}$  then, for all $v\in \Real^d$:
\begin{align*}
\mathscr D_W(z) =\inf \Big \{\Pi_z(w)-w.z:\prod_{k\in [d]}      (t_kw_k)^{{t_k}}=1, w\in \Real_{+}^d\Big \}.
\end{align*}
\end{prop}

\subsection{ Profit Function and Duality: Generalized Means and Directional Cases}\label{Dualp}
 In the following, we present a dual formulation of the generalized mean directional distance function in terms of the profit function. Specifically, we demonstrate that for \( p \geq 1 \), the generalized mean directional distance function can be interpreted as the maximization of a norm over a suitably restricted production set. By applying the maximum norm theorem, a duality result is derived. Following this approach, we propose a dual formulation of the generalized mean directional distance function and show that the dual properties of the asymmetric directional distance function emerge as a limiting case.

 {By definition we have
\begin{align}
\mathscr D_{W_{{p},g}}(z)=\left\{ \begin{matrix} \sup\limits_{v\in T_z} \Big(\sum\limits_{k\in
\mathcal G} \frac{1}{|\mathcal G|} \big(\frac{|u_k-z_k|}{g_k}\big)^p \Big)^{\frac{1}{p}} &\text{if }z\in T\\
+\infty &\text{otherwise.}\end{matrix}\right.
\end{align}
{Note that when $p > 0$, the proposed approach can be interpreted as a slack-based measure (see \cite{t01, fw09}). This interpretation no longer applies when $p \leq 0$, due to the absorption condition, which implies that if there exists some $i \in \mathcal G$ such that $u_i = 0$, then $W_{p,g}(u) = 0$. Therefore, a production vector whose distance to the frontier is zero may still be inefficient,  even though it is weakly efficient.  
In such cases, there remains potential for slack improvements.}  In the case where $p\geq 1$ one can provide a dual interpretation based upon the maximum norm theorem.  For example, if $g\in \Real_{++}^{d}$, then the map
\begin{equation} \|\cdot\|_{{ p, g^{-1}}}: \delta \mapsto
 \Big(  \frac{1}{ d}\sum_{k\in [d]}\big |\frac{\delta_k}{g_k} \big |^p\Big)^{\frac{1}{p}}\end{equation} defines a weighted norm on $\Real^{d}$ and we have for all $u\in \Real^d$ and all $p\in [1,\infty[$:
 \begin{equation} \|\delta\|_{{ p, g^{-1}}}=
 \Big( \sum_{k\in [d]}\big |\frac{1}{ d^{\frac{1}{p}}}\frac{\delta_k}{g_k} \big |^p\Big)^{\frac{1}{p}}.\end{equation}  If $p=\infty$ then
  \begin{equation}
   \|\delta\|_{{\infty,g^{-1}}}=\lim_{p\longrightarrow \infty}\|\delta\|_{ {p,g^{-1}}}=
      \max_{k\in [d]}\big |\frac{u_k}{g_k}\big|  .
  \end{equation}
  
 By definition, it follows that \begin{equation}\mathscr D_{W_{{p},g}}(z)=\sup\{\|u-z\|_{{p, g^{-1}} }: u\in T_z \}=\mathscr D_{\|\cdot \|_{{p,g^{-1} }}}(z).\end{equation}
   The dual norm is   defined by \begin{equation}\|w\|_{ q,g}=\Big( \sum_{k\in [d]}d^{\frac{q}{p}}|{g_k}{w_k}|^q\Big)^{\frac{1}{q}}=d^{\frac{q-1}{q}}\Big( \sum_{k\in [d]}|{g_k}{w_k}|^q\Big)^{\frac{1}{q}}\end{equation} with $\frac{1}{p}+\frac{1}{q}=1$,
where by definition $\|w\|_{{q,g}}=\sup_{}\{|\langle w,u\rangle |: \|u\|_{p,g^{-1},}=1\}$. In the above cases we have considered the situation where $g>0$. In the following we consider the general case. Let $\Pi_{z,g}: \Real_{+}^d \longrightarrow \Real\cup \{+\infty\}$ be the map defined as:
\begin{equation}
\Pi_{z,g}(w)=\sup_{u}\Big\{w.u:u\in T_z,  u_k=z_k, k\notin \mathcal G\Big\}.
\end{equation}
This profit function is limited to optimizing the netputs by fixing them at their initial levels when they are associated with a zero component of the direction $g.$
 
If $p<1$, we consider the quasi-concave utility functions:
\begin{equation}W_{{p},g}(\delta)=\stackrel{\phi_p}{\sum\limits_{k\in
[d]}}\frac{1}{d^{\frac{1}{p}}}\,\frac{\delta_k}{g_k} \;( p\not=0)\quad\text{and}\quad  W_{{0},g}(\delta)=\prod_{k\in[d]} \big(\frac{\delta_k}{g_k}\big)^{\frac{1}{d}}.\end{equation}
From equations \eqref{trans} and \eqref{transcobb}, the indirect utility functions are:

\begin{equation}W^{\star}_{p,g}(w)=\Big( \sum_{k\in [d]}d^{\frac{q-1}{q}}( {g_k}{w_k})^q\Big)^{-\frac{1}{q}}\; \quad\text{and}\quad 
W^{\star}_{0,g}(w)=   d\prod_{k\in [d]} \big( {g_k}  {w_k}\big)^{{-\frac{1}{ d}}}\end{equation}
 with $\frac{1}{p}+\frac{1}{q}=1$. The next results are obtained by applying our earlier results to the subspace $\Real_{\mathcal G}$ whose dimension is ${|\mathcal G|}$.

{ \begin{prop}\label{dual} Under $T1$  to $T$4, for all $z\in T$ and all $g\in \Real_+^d\backslash\{0\}$, we have the following properties:

\noindent $(a)$ For all $p\geq 1$ \begin{align*}\mathscr D_{W_{{p},g}}(z)= \sup\limits_{w\geq 0}& \Big \{\Pi_{z,g}(w)-w.z: |\small {\mathcal G}|^{\frac{q-1}{q}} \Big(  \sum_{k\in \mathcal G}({g_k}{w_k})^q\Big)^{\frac{1}{q}}=1\Big \}
\end{align*}  with $\frac{1}{p}+ \frac{1}{q}=1$.

\noindent $(b)$ If $p=1$, \begin{align*}\mathscr D_{W_{{1},g}}(z)= \mathrm{D}_{\mathrm{FL}}(z;g) =\sup\limits_{w\geq 0}  \Big \{\Pi_{z,g}(w)- w.z:  |\scriptsize {\mathcal G}|\max_{k\in \mathcal G} w_kg_k =1\Big\}.\end{align*}

\noindent $(c)$ If $p=\infty$

$$\mathscr D_{W_{\infty,g}}(z)=\mathrm{AD}(z;g)=\sup\limits_{w\geq 0} \big \{\Pi_{z,g}(w)- w.z:w.g=1 \big\}.$$

\noindent Suppose now that convexity holds $(T5)$, we have:\\

\noindent $(d)$ If $p=0$

\begin{align*} \mathscr D_{W_{{0},g}}(z) =\inf\limits_{w\geq 0}  \Big \{\Pi_{z}(w)-w.z: 
|\mathcal G| \prod_{k\in \mathcal G} \big( {g_k}  {w_k}\big)^{\frac{1}{|  \mathcal G|}}=1\Big\}.\end{align*}

\noindent $(e)$ If $p<1$ and $p\not=0$

\begin{align*} \mathscr D_{W_{{p},g}}(z) =\inf\limits_{w\geq  0}  \Big \{\Pi_{z}(w)- w.z:  |\mathcal G|^{\frac{q-1}{q}}\stackrel{\phi_q}{\sum_{k\in \mathcal G}} {g_k}{w_k}  =1\Big\}.\end{align*}

\noindent $(f)$ If $p=-\infty$
$$ \mathscr D_{W_{-\infty,g}}(z)=\mathrm{D}(z;g)=\inf\limits_{w\geq 0} \big \{\Pi_{z}(w)- w.z:w.g=1 \big\}.$$

\end{prop}

\begin{center}{\scriptsize \unitlength 0.4mm 
\linethickness{0.3pt}

\ifx\plotpoint\undefined\newsavebox{\plotpoint}\fi 

\ifx\plotpoint\undefined\newsavebox{\plotpoint}\fi 
\begin{picture}(305.5,159.25)(0,0)
\put(126,27.75){\vector(-1,0){121.75}}
\put(125.75,28.25){\vector(0,1){110.5}}
\put(163.5,28){\vector(-1,0){.07}}\multiput(288,27.75)(-15.5625,.03125){8}{\line(-1,0){15.5625}}
\put(288,28){\vector(0,1){108.5}}
\put(46.75,79.75){\circle*{1.414}}
\put(141.25,47.5){\vector(3,4){.07}}\multiput(125.75,28)(.0336956522,.0423913043){460}{\line(0,1){.0423913043}}
\put(305.5,46.75){\vector(1,1){.07}}\multiput(288.25,28.25)(.0336914063,.0361328125){512}{\line(0,1){.0361328125}}
\qbezier(287.5,28)(266.625,113.625)(166.25,138.75)
\put(192,76){\circle*{1}}
\put(0,27.75){\makebox(0,0)[]{$x$}}
\put(289.75,24.25){\makebox(0,0)[]{$0$}}
\put(130.75,23.25){\makebox(0,0)[]{$0$}}
\put(127.5,149){\makebox(0,0)[]{$y$}}
\put(288,144.25){\makebox(0,0)[]{$y$}}
\put(35.75,75.25){\makebox(0,0)[cc]{$z$}}
\put(187.25,69.75){\makebox(0,0)[cc]{$z$}}
\put(71.5,0){\makebox(0,0)[]
{{\bf Figure 12:} Duality in the case $p\in ]  1,+\infty[$}}
\put(222.25,0){\makebox(0,0)[]
{{\bf Figure 13:} Duality in the case $p\in ]  -\infty, 1[$}}
\put(91,60){\makebox(0,0)[]{$T$}}
\put(46.75,79.75){\line(0,1){59}}
\put(47.25,79.75){\line(1,0){67.25}}
\qbezier(107,62.25)(101.875,138.625)(20.25,143.5)
\put(82,84.75){\makebox(0,0)[]{$T_z$}}
\put(191.75,76.75){\line(0,1){64.5}}
\put(192,76.5){\line(1,0){84}}
\put(238.5,85.25){\makebox(0,0)[]{$T_z$}}
\put(248.25,58.75){\makebox(0,0)[]{$T$}}
\put(104.5,79.75){\circle*{1.803}}
\put(265.5,76.5){\circle*{1.5}}
\put(90.25,159){\makebox(0,0)[cc]{$\Pi_z(w)$}}
\put(164.25,159.25){\makebox(0,0)[cc]{$\Pi_z(w)$}}
\put(32.25,157){\makebox(0,0)[cc]{$w.z$}}
\qbezier(190.75,155.5)(236.625,87.625)(297,86.25)
\multiput(32.75,150)(.03370418848,-.16917539267){764}{\line(0,-1){.16917539267}}
\multiput(91.25,148.5)(.03370418848,-.16917539267){764}{\line(0,-1){.16917539267}}
\multiput(129.25,122.75)(.045658975362,-.033730934689){2557}{\line(1,0){.045658975362}}
\multiput(178.25,149.75)(.045658975362,-.033730934689){2557}{\line(1,0){.045658975362}}
\put(140,54){\makebox(0,0)[cc]{$g$}}
\put(303.5,54.75){\makebox(0,0)[cc]{$g$}}
\put(125.75,27.5){\line(0,-1){18.25}}
\put(287.75,27.5){\line(0,-1){18.25}}
\put(160.25,28.5){\makebox(0,0)[cc]{$x$}}
\put(145.75,115.5){\makebox(0,0)[cc]{$w.z$}}
\qbezier(126,27.25)(117.625,91.5)(66.75,97.75)
\qbezier(66.75,97.75)(35,101.5)(2.25,136.25)
\end{picture}

}
\end{center}

\noindent {The duality results established in Proposition \ref{dual}.(a), \ref{dual}.(b) \ref{dual}.(d)) have a  similar interpretation.
We consider, for any netput vector a suitable profit function restricted to the dominating production vectors.

The calculation of the generalized mean directional distance function reduces to determining a normalized dual price that minimizes the difference between the profit function and the value of a given netput. The differences among these measures arise from the normalization constraints imposed on the dual prices.

For \( p \in [1, \infty[ \), the computation of the generalized mean directional distance function involves maximizing a smooth \(\ell_p\) norm that identifies an efficient point on the frontier. In the directional F\"are-Lovell case, the normalization condition on shadow prices allows each netput price to be adjusted in coordinate directions to reach an efficient point. This adjustment contrasts with the directional distance function, where its additive structure enforces a projection along \( g \) for any netput vector.

The distinction between the asymmetric and standard directional cases, as established by the duality result in \cite{ccf98}, stems from their objective. In the former, the aim is to maximize the difference between the profit function and the profit evaluated at point \( z \), whereas in the latter, it is to minimize it. Furthermore, the shadow prices \( w^\star \), solutions to the dual problem, maximize the profit function at \( z^\star = z + \mathrm{D}(z;g)g \). As a result, \(\Pi_{z}(w^\star) = \Pi(w^\star)\), reaffirming the standard duality result.

More importantly, as established in Proposition \ref{limfunc}.(b),
\begin{equation}
\lim_{p \to -\infty} \mathscr D_{W_{{p},g}}(z) = \mathrm{D}(z;g).
\end{equation}
When \( p \to -\infty \), since \( q = \frac{p}{p-1} \), it follows that \( q \to 1 \). Under the normalization constraint in Proposition \ref{dual}.(e), we recover the standard normalization condition from \cite{ccf98}:
\begin{equation}
\lim_{q \to 1} \Big( | \mathcal G |^{\frac{q-1}{q}} \stackrel{\phi_q}{\sum_{k \in \mathcal G}} g_k w_k \Big) = w \cdot g = 1.
\end{equation}
{In addition, note that since $p \to 0_-$ implies $q \to 0_+$, it follows from the normalization condition in Proposition~\ref{dual}.(d) that
\begin{equation}
\lim_{q \to 0_+} \Big( |\mathcal G |^{\frac{q-1}{q}} \stackrel{\phi_q}{\sum_{k \in \mathcal G }} g_k w_k \Big)
= \lim_{q \to 0_+} |\mathcal G | \Big( \stackrel{\phi_q}{\sum_{k \in \mathcal G }} \frac{1}{|\mathcal G |} g_k w_k \Big)
= | \mathcal G | \prod_{k \in \mathcal{G}} (g_k w_k)^{\frac{1}{|\mathcal G |}}
= 1 .
\end{equation}
}

The formal differences in the duality results align with the taxonomy proposed in \cite{rs18}. This paper highlights structural differences between F\"are-Lovell and directional measures, corresponding to the cases \( p \geq 1 \) and \( p < 1 \), respectively.

Table 1 summarizes how to obtain the main directional distance function as limiting cases for different values of
the parameter $p$ in the  {generalized mean directional distance function}:\\
\begin{center}
{\scriptsize
\begin{tabular}{|c|c|}
\hline Value of $p$   & Distance function \\
\hline \hline
$-\infty$             & Directional Distance Function   \\

\hline $0$            & Multiplicative Directional F\"are-Lovell  \\
\hline $1$            & Directional F\"are-Lovell  \\
\hline $+\infty$      & Asymmetric Directional distance function  \\
\hline
\end{tabular}\\Table 1: Summary of Main Results}
\end{center}

{Table~1 shows that the proposed framework nests several existing approaches as limiting cases. In particular, the slack-based measures of \cite{fw09} that is an extenstion of the slack-based approach proposed in \cite{t01}, can be interpreted as nonparametric versions of the directional Färe--Lovell measure. Moreover, the asymmetric directional distance function introduced in \cite{ht02} was developed without an explicit notion of direction, which was later incorporated for input-oriented distance functions in \cite{bck11}. Note that a multiplicative directional distance function was also considered in \cite{msr14}. However, this efficiency measure is defined for a logarithmic technology within a nonparametric framework and is derived from a piecewise Cobb--Douglas production model.
 } Table 2 provides a taxonomy of the duality results relating the generalized mean directional F\"are-Lovell measure and the profit function for any value of $p$. In addition it indicates which dual optimization criterion is considered (maximisation or minimisation) in any case.

\begin{center}
{\scriptsize
\begin{tabular}{|c|c|c|c|c|}\hline Value of $p$   &  Optimization Criterion&Price Normalisation&Convexity\\
\hline \hline
$p=-\infty$ & Minimization&$ w.g=1$&Yes\\\hline
$p<0$&Minimization&$ |\mathcal G|^{\frac{q-1}{q}}\stackrel{\phi_q}{\sum\limits_{k\in \mathcal G}} {g_k}{w_k}  =1$&Yes\\
\hline $p=0$ & Minimization&$
{|\mathcal G|} \prod\limits_{k\in \mathcal G} \big( {g_k}  {w_k}\big)^{\frac{1}{|\mathcal G|}}=1$&Yes  \\\hline
$p\in ]0,1[$ &Minimization&$ |\mathcal G|^{\frac{q-1}{q}} \stackrel{\phi_q}{\sum\limits_{k\in \mathcal G}} {g_k}{w_k}  =1$&Yes\\
\hline $p=1$ & Maximization&$ \max\limits_{k\in \mathcal G}w_kg_k=1$&No\\\hline
$p>1$&Maximisation&$|\mathcal G|^{\frac{q-1}{q}}\Big( \sum\limits_{k\in \mathcal G}({g_k}{w_k})^q\Big)^{\frac{1}{q}}=1$&No\\
\hline $p=+\infty$ & Maximization&$w.g=1$&No\\
\hline
\end{tabular}\\Table 2: Duality Results}
\end{center}

{Table~2 presents two types of duality results. When $p < 1$, duality involves a minimization problem defined by the difference between the profit function and the profit associated with the evaluated production vector. The normalization of prices depends on the value of $p$, and the resulting formulation recovers the duality result established in \cite{ccf98}. In this case, the underlying function is quasi-concave, and duality is expressed in terms of an indirect utility function.  When $p \geq 1$, this difference is instead maximized. This change reflects the different nature of the efficiency measure, as the utility function becomes convex (indeed, a norm) and is maximized over a compact set.
In such a case, the problem under consideration is intrinsically non-convex, as it consists in maximizing a convex objective function over a compact feasible set,  even in cases where this set is convex. As a result, the associated dual problem is also more challenging to address from an operational and computational perspective.}

Finally, Table 3 summarizes the results in the norm and utility-based cases. 
\begin{center}
{\scriptsize
\begin{tabular}{|c|c|c|c|c|}\hline Utility-Based     &  Optimization Criterion&Price Normalisation & Convexity\\
\hline \hline
$\|\cdot \|$ & Maximization&$\|w\|^\star=1$& No\\\hline
$W$-Quasiconcave & Minimization&$W^\star(w)=1$& Yes\\\hline
\end{tabular}\\Table 3: Duality Results}
\end{center}

\section{Concluding Comments}
\label{sec:Concl}

We have introduced a new generalized mean directional distance function. This distance function extends the generalized F\"are-Lovell input efficiency measure recently proposed in \cite{bck20} to the full netput space. Building on the results established in \cite{bck20}, we have shown that the directional distance function \cite{ccf98} is recovered as \( p \to -\infty \), the F\"are-Lovell directional distance function \cite{bck11} is obtained when \( p = 1 \), the asymmetric directional distance function emerges as \( p \to -\infty \), and a directional version of the multiplicative F\"are-Lovell measure corresponds to \( p = 0 \).

It is worth noting that this article focuses only on cases where production vectors are feasible. Except for radial and directional measures, addressing the infeasible case presents significantly greater challenges.{ For example, the Färe-Lovell measure does not fully characterize the underlying production technology (see, for instance, \cite{s17}), which is an important property for the construction and interpretation of production indices or indicators.} Developing an approach to handle situations where these production vectors are not admissible would open the door to constructing a new class of productivity indicators .

\bigskip
\noindent {\bf Conflict of Interest Statement}

\noindent The author declares that there is no conflict of interest regarding the publication of this paper.\\

\noindent {\bf Funding}

\noindent This research did not receive any specific grant from funding agencies in the public, commercial, or not-for-profit sectors.

\section*{Appendix}
 
 \subsection*{Proofs of Propositions}
 \noindent {\bf Proof of Proposition \ref{UtForm}:} $(a)$  Suppose that $v\geq z$ with $v\not=z$.  Since $T_z$ is a compact subset of $\Real^d$, $W$ is upper semi-continuous and $T_v\subset T_z$, there exists $v^\star\in T_z$ such that
$W(v^\star-v)=\mathscr D_W (v)$. It follows that $v^\star-v\leq v^\star-z $ and $v^\star-v\not= v^\star-z $. Therefore, since $W$ is strongly increasing:
$\mathscr D_W (v)=W(v^\star-v)<W(v^\star-z)\leq \sup \{W(u-z):u\in T_z\}=\mathscr D_W (z)$. Hence $\mathscr D_W (v) < \mathscr D_W (z)$. Consequently, if $\mathscr D_W (z)=0$ then we cannot find any $u$ that strongly dominates $z$. Thus $z$ is efficient. To prove the reciprocal suppose that $\mathscr D_W (z)>0$. In such a case  there exists some $u\in T_z$ with $W(u-z)>0$. However, by hypothesis, if $u-z=0$, $W(u-z)=0$. Hence, we deduce that $u\not=z$ and this implies that $z\notin \mathcal E(T)$, which proves the reciprocal.  $(b)$ 
Suppose that $v> z$.  There exists $v^\star\in T_z$ such that
$W(v^\star-v)=\mathscr D_W (v)$. It follows that $v^\star-v< v^\star-z $. Therefore, since $W$ is weakly increasing:
$\mathscr D _W(v)=W(v^\star-v)<W(v^\star-z)\leq \sup \{W(u-z):u\in T_z\}=\mathscr D_W (z)$. Hence $\mathscr D_W (v) < \mathscr D_W (z)$. Consequently, if $\mathscr D_W (z)=0$ then we cannot find any $u$ such that   $z<u$ . Thus $z$ is weakly efficient. To prove the reciprocal suppose that $\mathscr D_W (z)>0$. In such a case  there exists some $u\in T_z$ with $W(u-z)>0$. However, from the absorption condition, if $u_k-z_k=0$ for some $k$, $W(u-z)=0$. Hence, we deduce that $u>z$ and this implies that $z\notin \mathcal F(T)$, which proves the reciprocal.    $(c)$ Let $z,v\in T$. Since $W$ is upper semi-continuous there exist $z^\star,v^\star\in T$ such that 
$\mathscr D_W (z)=W(z^\star-z)$ and $\mathscr D_W (v)=W(v^\star-v)$. This for all $\theta\in [0,1]$, $\theta z+(1-\theta)v\in T $ and $\theta z^\star +(1-\theta)v^\star\in T $. 
Since $W$ is quasi-concave
\begin{align*}W\big(\theta z^\star +(1-\theta)v^\star-(\theta z +(1-\theta)v )\big)&\geq \min\{W(z^\star-z),W(v^\star-v)\}\\&\geq \min\{\mathscr D_W (z),\mathscr D_W (v)\}.\end{align*}
However:
\begin{align*}W\big(\theta z^\star +(1-\theta)v^\star-(\theta z +(1-\theta)v )\big)&\leq \sup_{u\in  T_{\theta z+(1-\theta)v}}W(u-(\theta z +(1-\theta)v ))\\&\leq \mathscr D_W(\theta z+(1-\theta)v). \end{align*}
Hence $\mathscr D_W(\theta z+(1-\theta)v)\geq \min\{\mathscr D_W(z),\mathscr D_W(v)\}$ which ends the proof. 

$(d)$ Suppose that $D $ is not upper semi-continuous and let us show a contradiction. 
Let $z\in T$ and let us consider a sequence $\{z_n\}_{n\in \mathbb N}$ such that $\lim_{n\longrightarrow \infty}z_n=z.$ Suppose that 
$\lim\sup_{n\longrightarrow \infty}\mathscr D_W(z_n)>\mathscr D_W(z)$. This implies that there is an increasing subsequence $\{n_k\}_{k\in \mathbb N}$ such that 
$ \lim_{k\longrightarrow \infty}\mathscr D_W(z_{n_k})>\mathscr D_W (z)$. Since $W$ is upper semi-continuous, for any $k$ there exists $z_{n_k}^\star$ such that $W(z_{n_k}^\star-z_{n_k})=\mathscr D_W(z_{n_k})$. Suppose that $\underline z=\inf_{k}z_{n_k}$. By construction $z_{n_k}\in T_{\underline z}$ for every $n$. However $T_{\underline z}$ is a compact subset of $T$ and for all $n$, $z_{n_k}^\star\in T_{\underline z}$.  Therefore one can extract an increasing subsequence $\{  n_{k_q}\}_{q\in \mathbb N}$ such that 
$\{z_{n_{k_q}}^\star\}_{k\in \mathbb N}$ has a limit $z^\star\in T_{\underline z}$. Moreover, for any $q$, $z_{n_{k_q}}^\star\geq z_{n_{k_q}}$, therefore 
$z^\star\geq z$. Since $W$ is upper semi-continuous, it follows that:
$$ \lim_{q\longrightarrow \infty}(z_{n_{k_q}})=\lim_{q\longrightarrow \infty}W(z_{n_{k_q}}^\star-z_{n_{k_q}})\leq \limsup_{n\longrightarrow \infty}W(z_{n }^\star-z_{n }) \leq W(z^\star-z)\leq \sup_{u\in T_z}W(u-z)\leq \mathscr D_W (z).$$
However, this is a contradiction,  which proves  $(d)$. $\Box$\\

\noindent{\bf Proof of Proposition \ref{LimGen}:}  Since the convergence is uniform over any compact subset $ K$ of $\Real_+^d$,   $W$ is continuous over $\Real_+^d$. Therefore,  there is some $u^\star\in T_z$ such that $W(u^\star-z)=\sup_{u\in T_z}W(u-z). $ Since $\lim_{p\longrightarrow \infty}W_{{p}}(u^\star-z)=W(u^\star-z)$, it follows that 
$$ \liminf_{p\longrightarrow  \infty } \mathscr D_{W_{{p}}}(z)=\liminf_{p\longrightarrow  \infty } \sup_{u\in T_z}W_{{p}}(u-z)\geq   \liminf_{p\longrightarrow \infty}W_{{p}}(u^\star-z) = \sup_{u\in T_z}W (u-z). $$
Conversely, for each $p$, since $W_{{p}}$ is continuous 
 this implies that  there is some $u_{{p}}\in T_z$ such that $\sup_{u\in T_z}W_{{p}} (u-z) = W_{{p}} (u_{{p}}-z)$. Therefore 
 $$\limsup_{p\longrightarrow  \infty } \mathscr D_{W_{{p}}}(z)=\limsup_{p\longrightarrow  \infty } \sup_{u\in T_z}W_{{p}}(u-z)=\limsup_{p\longrightarrow \infty }W_{{p}  }(u_{{p}}-z).$$ Since $T_z$ is a compact subset of $\Real^d$, it follows that there exists an increasing sequence $\{p_k\}_{k\in \mathbb N}$ and some $\bar u \in T_z$ such that $\lim_{k\longrightarrow +\infty}u_{{p_k}}^\star=\bar u  $ and
 $$\limsup_{p\longrightarrow  \infty } \mathscr D_{W_{{p}}}(z)=\limsup_{p\longrightarrow \infty }W_{{p }}(u_{{p}}-z)=\lim_{k\longrightarrow \infty }W_{{p_k}}(u_{{p_k}}-z).$$
 
  However, using the triangular inequality, we have:
 \begin{align*}|W_{{p_k}}(u_{{p_k}}-z)-W(\bar u -z)|&\leq |W_{{p_k}}(u_{{p_k}}-z)-W(u_{{p_k}}-z)|+| W(u_{{p_k}}-z)-W(\bar u -z)|\\&\leq  \sup_{u\in T_z}|W_{{p_k}}(u-z)-W(u-z)|+| W(u_{{p_k}}-z)-W(\bar u-z)|. \end{align*}
 Consequently, $\limsup_{p\longrightarrow  \infty } \mathscr D_{W_{{p}}}(z)=\lim_{k\longrightarrow \infty }W_{{p_k}}(u_{{p_k}}-z)=W(\bar u-z)\leq \sup_{u\in T_z}W (u-z)$, which implies that 
 $$ \limsup_{p\longrightarrow  \infty } \mathscr D_{W_{{p}}}(z)\leq \sup_{u\in T_z}W (u-z)\leq \liminf_{p\longrightarrow  \infty } \mathscr D_{W_{{p}}}(z).$$
 Finally, conditions $(i)$ and $(ii)$ in Definition \ref{DefUbased} are satisfied from the continuity of $W$. Since for all $p$, $W_{{p}}$ is non-decreasing condition $(iii)$ immediately follows.  $\Box$\\
 
 \noindent{\bf Proof of Corollary \ref{CorLimGen}:} Since $\{W_{{p}}\}_{p\in \mathbb N}$ is a  monotonic sequence of functions, from the Dini's theorem, we deduce that $\{W_{{p}}\}_{p\in \mathbb N}$ uniformly converges to $W$ for any compact subset $ K$ of $\Real^d. $  Hence, the result follows from  Proposition \ref{LimGen}. $\Box$\\

\noindent  {\bf Proof of Proposition \ref{monotoneNew}:} First we prove that if $\delta\in \Real_+^n$ and then  the map $p\mapsto \big(\frac{1}{d}\sum_{i\in [d]}\delta_i^p\big)^{\frac{1}{p}}$ is increasing in $p $ (see also \cite{b03}}. Suppose that $p>q>0$ and let us consider the map $\phi_{\frac{q}{p}}$. Since $q<p$ it follows that $\phi_{\frac{q}{p}}$ is concave.   From the concavity of $\phi_{\frac{q}{p}}$ over $\Real_+^d$:
$$\phi_{\frac{q}{p}}\Big(\frac{1}{d}\sum_{i\in [d]} {\delta_i}^{p}\Big)\geq \frac{1}{d}\sum_{i\in [d]}\phi_{\frac{q}{p}}({\delta_i}^p)=\frac{1}{d}\sum_{i\in [d]} {\delta_i}^q.$$
 Now, since the map $\phi_{\frac{1}{q}}$ is increasing and since $\phi_{\frac{1}{q}}\circ \phi_{\frac{q}{p}}=\phi_{\frac{1}{p}}$ we deduce that
$$\big(\sum_{i\in [d]}\frac{1}{d}\delta_i^p\big)^{\frac{1 }{p}}\geq  \big(\sum_{i\in [d]}\frac{1}{d}\delta_i^q \big)^{\frac{1 }{q}}.  $$
Suppose now that $0>p>q$. If $\delta_i=0$ for some $i$, $ \stackrel{\phi_p}{\sum\limits_{i\in [d]}}\frac{1}{d^{\frac{1}{p}}}\delta_i=0$ and the result is immediate. Suppose that 
$\delta_i>0$ for all $i$. Since $p>q$ and $\frac{p}{q}=|\frac{p}{q}|<1$, we deduce that:
$$\phi_{\frac{p}{q}}\Big(\frac{1}{d}\sum_{i\in [d]} {\delta_i}^{q}\Big)\geq \frac{1}{d}\sum_{i\in [d]}\phi_{\frac{p}{q}}({\delta_i}^q)=\frac{1}{d}\sum_{i\in [d]} {\delta_i}^p .$$
Since the map $\lambda \mapsto \lambda ^{\frac{1}{p}}$ is decreasing:
$$ \Big(\frac{1}{d}\sum_{i\in [d]} {\delta_i}^{q}\Big)^{\frac{1}{q}}\leq \Big(\frac{1}{d}\sum_{i\in [d]} {\delta_i}^p \Big)^{\frac{1}{p}},$$
which ends the first part of the proof. 
Now since for all $\delta \in \Real_{++}^d$,  $\lim_{p\longrightarrow 0_-}\stackrel{\phi_p}{\sum\limits_{i\in [d]}}\frac{1}{d^{\frac{1}{p}}}\delta_i=\prod\limits_{i\in [d]}\delta_i^{\frac{1}{d}}$, we deduce that for all $q<0$, $ \stackrel{\phi_q}{\sum\limits_{i\in [d]}}\frac{1}{d^{\frac{1}{q}}}\delta_i\leq \prod\limits_{i\in [d]}\delta_i^{\frac{1}{d}} $. Finally, from the concavity of the logarithm function, we deduce that for all $p>0$, and all $\delta\in \Real_{++}^d$, $\prod\limits_{i\in [d]}\delta_i^{\frac{1}{d}}\leq  \stackrel{\phi_p}{\sum\limits_{i\in [d]}}\frac{1}{d^{\frac{1}{p}}}\delta_i $ since $\prod\limits_{i\in [d]}\delta_i^{\frac{1}{d}}=0$ if there is some $\delta_i=0$, we deduce that for all $p,q$, if $p\geq q$, then  $\frac{1}{d^{\frac{1}{q}}}\stackrel{\phi_q}{\sum\limits_{i\in [d]}}\delta_i\leq  \frac{1}{d^{\frac{1}{p}}}\stackrel{\phi_p}{\sum\limits_{i\in [d]}}\delta_i$. Now, by construction we have:
$$\lim_{ k\longrightarrow \infty  }   W_{{p_k}} =  W_{{\bar p}}.$$
Since  $\{p_k\}_{k\in \mathbb N}$ is a monotone sequence, the sequence $\{W_{{p_k}}\}_{k\in \mathbb N}$ is monotone and it follows from Corollary \ref{CorLimGen} that:
$$\lim_{ k\longrightarrow \infty  }\mathscr D_{W_{{p_k}}}=\mathscr D_{W_{{\bar p}}}. \quad \Box$$

\noindent {\bf Proof of Proposition \ref{maxfar}:}  $(a)$ Note that the map $\delta\mapsto \stackrel{\phi_p }{\sum\limits_{k\in [d]}} \delta_k$ is non-decreasing on $\Real^d  $.
One can equivalently write:
\begin{align*}
\delta_{{p}}:= \sup_{\delta\geq 0} \Big \{ \frac{1}{{|\mathcal G|}^{\frac{1}{p}}} \stackrel{\phi_p}{\sum\limits_{k\in
 \mathcal G}} \delta_k :  z+\delta\odot g\leq u, u\in T_z \Big\}= \sup_{\delta\geq 0} \Big \{\frac{1}{{|\mathcal G|}^{\frac{1}{p}}} \stackrel{\phi_p}{\sum\limits_{k\in
 \mathcal G}} \delta_k :    \delta\odot g\leq u-z, u\in T_z \Big\}.
\end{align*}
Therefore:
\begin{align*}
 \delta_{{p}}:= \sup_{\delta\geq 0} \Big \{  \frac{1}{{|\mathcal G|}^{\frac{1}{p}}} \stackrel{\phi_p}{\sum\limits_{k\in
 \mathcal G}} \delta_k :     \delta_k \leq \frac{u_k-z_k}{g_k}, k\in \mathcal G,  u\in T_z \Big\}=\sup_{u}\Big \{W_{p,g}(u-z) :   u\in T_z \Big\}=\mathscr D_{W_{{p},g}}(z),
\end{align*}
which yields the result.  The proof of $(b)$  is obtained similarly. $(c)$  Equivalently we can write:
\begin{equation*}
\mathrm{D}(z;g)= \sup_{\delta}\Big \{\delta :   z+\delta g\leq   u\in T_z \Big\}=\sup_{\delta}\Big \{\delta :   \delta g\leq u-z, u\in T_z \Big\}.
\end{equation*}
Hence
\begin{equation*}
\mathrm{D}(z;g)= \sup_{\delta}\Big \{\delta :  \delta \leq \min_{k\in \mathcal G}\frac{u_k-z_k}{g_k}, u\in T_z \Big\}=\sup_{u}\Big \{W_{-\infty,g}(u-z) :   u\in T_z \Big\}=\mathscr D_{W_{-\infty,g}}(z).
\end{equation*}
 
$(d)$ By definition, we have:
\begin{equation*}
\mathrm{AD}(z;g)= \max_{k\in \mathcal G}\sup_{ u\in T_{z}}    \big\{z+\delta g_ke_k\leq u\big\}. 
\end{equation*}
Since $T2$ holds, we have:
\begin{equation*}
\mathrm{AD}(z;g)= \max_{k\in \mathcal G}\sup_{ u\in T_{z}}    \big\{z_k+\delta g_k\leq u_k\big\}. 
\end{equation*}
We deduce that
\begin{equation*}
\mathrm{AD}(z;g)= \max_{k\in \mathcal G} \sup_{ u\in T_{z}}   \big\{\frac{u_k-z_k}{g_k}\big\}= \sup_{ u\in T_{z}} \max_{k\in \mathcal G}   \big\{\frac{u_k-z_k}{g_k}\big\}=\mathscr D_{W_{\infty,g}}(z). \Box
\end{equation*}

\noindent {\bf Proof of Lemma \ref{compact}:} The proof immediately follows from Proposition \ref{maxfar} and from the fact that for all $p\in [-\infty,+\infty]$, $W_{{p},g}$ is upper semi-continuous.  $\Box$\\

\noindent {\bf Proof of Proposition \ref{AXPROP}:} The proofs of $(a)-(a')$ and $(c)-(c')$ are similar to the proofs of  $(a)$ and $(b)$ in Proposition \ref{UtForm}. The proofs of $(b)$ and $(d)$ are immediate. $\Box$\\

\noindent{\bf Proof of Proposition \ref{limfunc}:}   We have for all $u\in T_z$:
$$\lim_{p\longrightarrow -\infty}W_{{p},g}(u-z)= W_{-\infty,g}(u-z)=\min_{k\in \mathcal G}\Big\{\frac{u_k-z_k}{g_k}\Big\};$$
 $$\lim_{p\longrightarrow 0^-}W_{{p},g}(u-z)= W_{{0},g}(u-z)=\prod_{k\in \mathcal G}\big(\frac{u_k-z_k}{g_k}\big)^{\frac{1}{{|\mathcal G|}}};$$
 $$\lim_{p\longrightarrow +\infty}W_{{p},g}(u-z)= W_{\infty,g}(u-z)=\max_{k\in \mathcal G}\Big\{\frac{u_k-z_k}{g_k}\Big\}.$$

For any monotone sequence of real numbers $\{p_k\}_{k\in \mathbb N}$ and all $u\in T_z$, 
the sequence $\{W_{{p_k},g}(u-z)\}_{k\in \mathbb N}$ is monotone. Suppose that $\{p_k\}_{k\in \mathbb N}$  is monotone and converges to some $\bar p\in \Real\cup\{-\infty, +\infty\}$. Since map $W_{{0}}$, $W_{-\infty}$ and $W_{\infty}$ are continuous, the convergence is uniform and 
$$\lim_{k\longrightarrow \infty}\mathscr D_{W_{{p_k},g}}(z)=\lim_{k\longrightarrow \infty}\sup_{u\in T_z}W_{{p_k},g}(u-z)=\sup_{u\in T_z}W_{{\bar p},g}(u-z)=\mathscr D_{W_{{\bar p},g}}(z).$$ Since this is true for every monotone sequences  $\{p_k\}_{k\in \mathbb N}$, we deduce the result. $\Box$\\

\noindent {\bf Proof of Proposition \ref{Order}:} For all real numbers $p,q$, if $p\geq q$ then $W_{{p},g }\geq W_{{p},g }$ which implies that 
$\mathscr D_{W_{{q},g}}(z) \leq \mathscr D_{W_{{p },g}}(z)$. Since 
$$\lim_{p\longrightarrow -\infty}\mathscr D_{W_{{p },g}}(z)=\mathrm{D}(z;g)\quad \text{ and }\quad \lim_{p\longrightarrow +\infty}\mathscr D_{W_{{p},g}}(z)=\mathrm{AD}(z;g)$$
we deduce the result. $\Box$\\

\noindent{\bf Proof of Proposition \ref{normprod}}: From Proposition \ref{maxnorm}, we have:
$$\max \{\|u-z\|: u\in T_z\}=\sup_{w\in \Real^d} \{h_{T_z}(w)- w.z: \|w\|^\star=1\}. $$ Moreover, there exists a point $u^\star\in \Real^d$ such that
$\|u^\star-z\|=\mathscr D_{\|\cdot\|}(z)$. Moreover, there exists some $w^\star\in \Real^d$ such that:
$$w^\star. u^\star=h_{T_z}(w^\star).$$ However, since $\|\cdot \|$ is weakly increasing over $\Real_+^d$, it follows that $T_z\cap \big(u^\star +\Real_{++}^d\big)=\emptyset$. 
Consequently $\{u\in \Real^d: w^\star.u=h_{T_t}(w^\star)\}\cap \Real_{++}^d=\emptyset$. From the Farkas Lemma, this implies that $w^\star\geq 0$. Therefore:
$$h_{T_z}(w)=\Pi_z(w^\star)$$
which proves the first par of the result. The second part immediately follows. $\Box$\\

\begin{lem} \label{maxlem}Let $a,b\in \Real_{++}^d$ and $c> 0$ be a nonnegative real number. \\
$(a)$ For all $p<1$ with $p\not=0$, let us consider the problem
\begin{equation}
\sup_{v\in \Real_+^d}\Big \{\stackrel{\phi_p}{\sum_{k\in [d]}}a_k v_k  :  \langle b, v\rangle =c,v\in \Real_+^d\Big \}.
\end{equation}
Then there is a solution $v^\star\in \Real_{++}^d$ with
\begin{equation*}
 v_k^\star={c}\Big[\sum_{k\in [d]} ({a_k}^{-1} {b_k})^{\frac{ p}{ p-1}}\Big]^{-1}{({a_k}^{-p}b_k)}^{ \frac{1}{ p-1}}.
\end{equation*}
Moreover
\begin{equation*}
\big(\sum_{k\in [d]} {a_kv_k^\star}^p\big)^{\frac{1}{p}}=
 {c}\big(\sum_{k\in [d]} ({{a_k^{-1}} {b_k})}^{ q}\big)^{-\frac{1}{q}},
\end{equation*}
where $\frac{1}{p}+\frac{1}{q}=1$.

\noindent $(b)$ Suppose that $t\in \Real_{++}^d$ and let us consider the problem
\begin{equation*}
\sup_{}\Big \{ \prod_k ( {v_k})^{t_k}:  \langle b, v\rangle =c, v\in \Real_+^d\Big \}.
\end{equation*}
Then there is a solution $v^\star$ with $ v_k^\star=c {t_k}  \Big(\sum\limits_{k\in [d]}  {  t_k } \Big)^{-1} {b_k}^{-1}$.
The value of the objective function is then:
\begin{equation*}
 \prod_{k\in [d]} (t_k {v_k^\star})^{t_k} = \prod_{k\in [d]} \Big[c  {t_k}  \Big(\sum_{k\in [d]}  {  t_k } \Big)^{-1} {b_k}^{-1}\Big]^{{t_k}}.
\end{equation*}

\end{lem}

\noindent {\bf Proof of Lemma  \ref{maxlem}:} $(a)$ Since $a,b\in \Real_{++}^d$ and $c>0$, one can find some $v\in \Real_{++}^d$ such that $\stackrel{\phi_p}{\sum_{k\in [m]}}a_k v_k>0$. Hence, since $v_k=0$ for some $k$ implies that $\stackrel{\phi_p}{\sum_{k\in [m]}}a_k v_k=0$, any solution of the optimization problem belongs to  $\Real_{++}^d$. Therefore,   the problem can equivalently be written:
$$
\sup_{v}\Big\{\ {\sum_{k\in [d]}}{(a_k v_k)}^p:  \langle b, v\rangle =c\Big\}.
$$
The objective function is quasi-concave and the Lagrangian is defined as:
$$
L(v,\lambda)= {\sum_{k\in [d]}}{(a_k v_k)}^{p}-\lambda  (\langle b, v\rangle -c).
$$
The first order condition yields:
$$
\frac{\partial L(v,\lambda)}{\partial x_k}=  {a_k}^p\frac{p}{{v_k}^{1-p}}-\lambda b_k=0.
$$
Therefore, for all $k$
$$
 v_k=\Big(\frac{p }{\lambda b_k}\Big)^{\frac{1}{1-p}}{a_k}^{\frac{p}{1-p}}.
$$
Hence, from $\frac{\partial L(v,\lambda)}{\partial \lambda}=0$, we deduce that
$$
 \langle b,v\rangle =\sum_{k\in [d]}b_kv_k=\sum_{k\in [d]}b_k\Big(\frac{p }{\lambda b_k}\Big)^{\frac{1}{1-p}}{a_k}^{\frac{p}{1-p}}=
 \sum_{k\in [d]} \Big(\frac{p{a_k}^p {b_k}^{-p}}{\lambda}\Big)^{\frac{1}{1-p}}=c.
$$
Hence
$$\Big(\frac{ p}{\lambda}\Big)^{\frac{1}{1-p}}\sum_{k\in [d]} \big({a_k}^{-1} {b_k}\big)^{\frac{-p}{1-p}}=c
\quad \text{and}\quad
 \lambda=p
 \Big(\frac{1}{c}\sum_{k\in [d]} \big( {a_k}^{-1}{b_k}\big)^{\frac{-p}{1-p}}\Big)^{ {1-p}}.
$$
Therefore
for all $i$
$$ v_k=\Bigg(\frac{p{a_k}^p}{p
 \Big(\frac{1}{\alpha}\sum_{k\in [d]} \big( {a_k}^{-1}{b_k}\big)^{\frac{-p}{1-p}}\Big)^{ {1-p}} y_k}\Bigg)^{\frac{1}{1-p}}=
 {c}\Big[\sum_{k\in [d]} ({a_k}^{-1} {b_k})^{\frac{-p}{1-p}}\Big]^{-1}{({a_k}^{p}{b_k}^{-1})}^{ \frac{1}{1-p}}.
$$
The value of the objective function is then:
\begin{align*}
(\sum_{k\in [d]} {(a_kv_k)}^p)^{\frac{1}{p}}&=
 {c}\Big[\sum_{k\in [d]} ({a_k}^{-1} {b_k})^{\frac{-p}{1-p}}\Big]^{-1}\big(\sum_{k\in [d]}{a_k}^{p}({{a_k}^{p}{b_k}^{-1})}^{ \frac{p}{1-p}}\big)^{\frac{1}{p}}\\&=
 {c}\Big[\sum_{k\in [d]} ({a_k} {b_k}^{-1})^{\frac{p}{1-p}}\Big]^{-1}\big(\sum_{k\in [d]} ({{a_k} {b_k}^{-1})}^{ \frac{p}{1-p}}\big)^{\frac{1}{p}}\\&=
 {c}\big(\sum_{k\in [d]} ({{a_k^{-1}} {b_k})}^{ q}\big)^{-\frac{1}{q}} .
\end{align*}
$(b)$ The result is standard and the proof is therefore omitted. $\Box$\\

\noindent {\bf Proof of Lemma \ref{strictpos}} First note that since $C$ is a comprehensive and convex subset of $\Real_+^d$ that contains $0$:
$$C=\bigcap\limits_{w\in \Real_{+}^d}\{z:w.x\leq h_C(w)\}.$$Suppose that there is some $z\in \Real_+^d$ such that $z\notin C$ and $z\in \bigcap\limits_{w\in \Real_{++}^d}\{z:w.x\leq h_C(w)\}$. Suppose that  $z\notin C$, and let us prove a contradiction. In such a case, there is some $\bar w\in \Real_+^{d}$ such that
$\bar w.z>h(\bar w)$. Let $\mathcal K(\bar w)=\{k\in [d]: \bar w_k>0\}$. Let $\{w^{(n)}\}_{n\geq 1}$ be a sequence of positive vectors such that:
$$w_k^{(n)}=\left\{\begin{matrix}
\bar w_k&\text{if}& k\in \mathcal K(\bar w)\\
\frac{1}{n}&\text{if}& k\notin \mathcal K(\bar w). 
\end{matrix}\right.$$
By hypothesis, for all $n$, $w^{(n)}.z \leq h_C(w_n)$ and $\bar w .z >h_C(\bar w )$. 
However, the latter condition implies that there is some $\epsilon >0$:
$$h(\bar w)+\epsilon<\bar w.z=\sum_{k\in \mathcal K(\bar w)}\bar w_kz_k\leq \sum_{k\in \mathcal K(\bar w)}\bar w_kz_k+\frac{1}{n} \sum_{k\notin \mathcal K(\bar w)}z_k\leq h(   w^{(n)}).$$
Now, note that since $C$ is compact there is a nonnegative  $M$ such that $\sup \big\{\sum_{k\in [d]}z_z:z\in C\big\}\leq M$. Moreover
\begin{align*}h(   w^{(n)})&= \sup_{u\in C}\big \{\sum_{k\in \mathcal K(\bar w)}\bar w_ku_k+\frac{1}{n} \sum_{k\notin \mathcal K(\bar w)}u_k\big\}\\&\leq  \sup_{u\in C} \big\{\sum_{k\in \mathcal K(\bar w)}\bar w_ku_k\big\}+\frac{1}{n} \sup_{u\in C}\big\{\sum_{k\notin \mathcal K(\bar w)}u_k\big\}\leq h_C(\bar w)+\frac{M}{n}\end{align*}
 Therefore 
 $$h(\bar w)+\epsilon\leq h_C(\bar w)+\frac{M}{n}.$$
However, there exists some $n$ sufficiently large such that $\frac{M}{n}<\epsilon$ that is a contradiction. Consequently $z\in C$, which ends the proof.
$\Box$\\

\noindent {\bf Proof of Proposition \ref{quasiconcdual}:}   For all $z\in \Real_+^d$, since $T_z$ is a compact set, there is some $\alpha\geq 0$ such that:
$$\sup_{u}\Big \{W(u-z) :  u\in T_z \Big \}=\alpha.$$
It follows that for all $\epsilon>0$:
$$ \Big \{u:W(u-z) >\alpha+\epsilon  \Big \}\cap T_z=\emptyset.$$
 Since $T_z$ satisfies $T_2$ and $W$ is upper-semicontinuous and nondecreasing map, it follows that
$$\sup_{u}\Big \{W(u-z) : u\in T_z \Big \} =\sup_{u}\Big \{W(u-z) :  u\in (T_z-\Real_+^d)\cap \Real_+^d\Big \}.  $$
Moreover
$$ T_z=\bigcap_{w\in \Real_{+ }^d}\{u\in \Real_+^d:   w .  u \leq \Pi_z(w)\}. $$
From the compactness of $T_z$, for all non-negative vector  $u'\notin T_z$, 
there exists some $w'\in \Real_{++}^d$ such that $  w'. u'  > \Pi_z(w')$. It follows that
$$ T_z=\bigcap_{w\in \Real_{++}^d}\{u\in \Real_+^d:   w .  u \leq \Pi_z(w)\}. $$
Since the map $W$ is quasi-concave, it follows that $\{u:W(u-z) \geq\alpha+\epsilon\} $ is convex and from the upper-continuity of 
$W$, it is closed. Hence for every $\epsilon>0$, there exists some $w_\epsilon\in \Real_{++}^d$ such that
$$ \{u:W(u-z) >\alpha+\epsilon\}\subset \{u:   w_\epsilon . u  \geq \Pi_z(w_\epsilon)\}$$
and
$$ T_z\subset \{u:   w_\epsilon \cdot  u \rangle\leq \Pi_z(w_\epsilon)\}.$$
Consequently,
$$\sup_{u}\Big \{W(u-z) :  w_\epsilon. u \leq \Pi_z(w_\epsilon)  \Big \}\leq\alpha+\epsilon.$$
Since this is true for every $\epsilon>0$, we deduce from Lemma \ref{maxlem} that:
\begin{align*}\sup_{u}\Big \{W(u-z):  u\in T_z  \Big \}&= \inf_{w\in \Real_{++}^d}\sup_{u} 
\{W(u-z):   w . u  \leq \Pi_z(w)\Big\}\\&=\inf_{w\in \Real_{++}^d}\sup_{v} \Big\{W(v):   w \cdot v = \Pi_z(w)-w. z\Big\}\\&=\inf_{w\in \Real_{++}^d}\Big\{ ( \Pi_z(w)-w.z)W^\star(w)\Big\}\\&=\inf_{w\in \Real_{++}^d}\Big\{{ \Pi_z(w)-w.z}:W^\star(w)=1\Big\}.
\end{align*}
Now suppose that $w$ is a normalized vector in $\Real_+^d$  that may have some null components. Since $T_z\leq \{u: w.z\leq \Pi_z(w)\}$ we deduce that
$\mathscr D_W(z)\leq \sup_u\{W(u-z): w.u\leq \Pi_z(w)\}=(\Pi_z(w)-w.z)W^\star(w)=\Pi_z(w)-w.z$. Therefore:
\begin{align*}\mathscr D_W(z)=\inf_{w\in \Real_{+}^d}\Big\{{ \Pi_z(w)-w.z}:W^\star(w)=1\Big\}. \quad\Box
\end{align*}

\noindent {\bf Proof of Proposition \ref{dual}:} $(a)$ Let us denote $\Real_{\mathcal G}= \big\{\sum_{k\in \mathcal G}z_ke_{k}: z_k\in \Real, k\in \mathcal G\big\}$. $\Real_{ \mathcal G} $ is a vector subspace of $\Real^{d}$ of dimension ${|\mathcal G|}$. The norm \begin{equation} \|\cdot\|_{{p,g^{-1}} }: u\mapsto
 \Big(\sum_{k\in \mathcal G}   \frac{1}{ |\mathcal G|}\sum_{k\in \mathcal G}\big |\frac{u_k}{g_k} \big |^p\Big)^{\frac{1}{p}}\end{equation} defines a weighted norm on $\Real_{\mathcal G }$ and a topology on this subspace. The dual norm is then defined by \begin{equation}\|u\|_{ q,g}=\Big(  \sum_{k\in \mathcal G}{|\mathcal G|}^{\frac{q}{p}}|{g_k}{u_k}|^q\Big)^{\frac{1}{q}}\end{equation} with $\frac{1}{p}+\frac{1}{q}=1$. We have shown that:
\begin{align*}
\mathscr D_{W_{{p},g}}(z)= \sup_{u\geq z}&\Big \{  \stackrel{\phi_p}{\sum\limits_{k\in
 \mathcal   G}} \frac{u_k-z_k}{g_k} : u\in T \Big\}.
\end{align*}
For the sake of simplicity, let us denote $z'=\sum_{k\in \mathcal G}z_ke_k$ and $u'=\sum_{k\in \mathcal G}u_ke_k$ the respective projections of $z$ and $u$ onto $\Real_{\mathcal G}$.  Moreover, let 
$$T_z'=\{\sum_{k\in \mathcal G}u_ke_k: u\in T_z\}$$
 be the  projection  of   $T_z$ onto $\Real_{\mathcal  G}$. By hypothesis $u\geq z$ implies that $ u'\geq  z'$.   Since 
$T$ satisfies $T_2$, if $u\geq z$ and $u\in T$, then the vector $v=u'+(z-z')  $ is in $T_z$. Hence
$v-z=u'-z'\in \Real_{\mathcal G}$. This implies that $v_k=z_k$ for all $k\notin \mathcal G$. Furthermore,   $u'\in T'_z$ since $T_2$ holds:
\begin{align*}\sup_{u\geq z}\Big \{  \stackrel{\phi_p}{\sum\limits_{k\in
 \mathcal G}} \frac{u_k-z_k}{g_k} : u\in T \Big\}&=\sup_{v\geq z}\Big \{  \stackrel{\phi_p}{\sum\limits_{k\in
\mathcal G}} \frac{v_k-z_k}{g_k} : u\in T \Big\}\\&= \sup_{u' }\Big \{\| z'-u' \|_{{p,g^{-1}}} : u'\in  T'_z \Big\}.\end{align*}
Equivalently:
\begin{align*}
\mathscr D_{W_{{p} ,g}}(z)= \sup_{u' }\Big \{d_{p, g^{-1}}\big(u', z'\big) : u'\in  T'_z \Big\}.
\end{align*}
For all $w'\in \Real_{ \mathcal G}$ let us denote
$$h'_{T'_z} (w')=\sup_{v'}\{\langle w',v'\rangle :v'\in T'_z\}. $$
$h'_{T'_z}$ is the functional support of $T'_z$ with respect to $\Real_{\mathcal  G}$. From Proposition \ref{maxnorm}:
\begin{align*}\max  \{d_{ p, g^{-1}}\big(u', z'\big):  u'\in T'_z\} =\sup_{w'\in \Real_{\mathcal  G }}
\{h_{T'_z} (w') - \langle w',z'\rangle: \|w'\|_{ q,g}=1\}, \end{align*}
However, for all $w,u, z\in \Real^d$
$$h'_{T'_z} (w')- \langle w',z'\rangle =\Pi_{z,g}(w)-w.z\quad \text{and}\quad  \|w'\|_{q,g}=|\mathcal G|^{\frac{q-1}{q}}\Big(  \sum_{k\in \mathcal G} |{g_k}{w_k}|^q\Big)^{\frac{1}{q}}. $$
We deduce $(a)$. 

\noindent $(b)$ If $p=1$ then $$\|w\|_{{{1,g^{-1} }}}=  \frac{1}{ |\mathcal G|}\sum_{k\in \mathcal G}|u_k| .$$ The dual norm is
$$\|w\|_{{\infty, g}}=  |\mathcal G|\max_{k\in \mathcal G}|g_k w_k| .$$

\noindent $(c)$ If $p=\infty$, we have
$$\|u\|_{{\infty,g^{-1}} }=  \max_{k\in \mathcal G}\big |\frac{u_k}{  g_k} \big | .$$
Since $q=1$, the dual norm is
$$\|u\|_{ 1,g}=  \sum_{k\in \mathcal G}{}|g_ku_k| .$$
 From Proposition \ref{maxnorm}, we deduce the result. $(d)$ We have
  $$\mathscr D_{W_{{0}}}(z)=\sup_{u\in T_z} \prod_{k\in \mathcal G}\big(\frac{u_k-z_k}{g_k}\big)^{\frac{1}{|\mathcal G|}}.$$
The set $\{ \sum_{k\in \mathcal G}(u_k-z_k)e_k:u\in T_z\}$ is a convex and comprehensive subset of $\Real_{\mathcal  G,+ }$. Therefore, we deduce the result applying Proposition   \ref{dualneg}.(b). We have, using the notations of  $(a)$
\begin{align*} \mathscr D_{W_{{0}}}(z) =\inf\limits_{w'\geq 0}  \Big \{\Pi_{z,g}(w')-w'.z': 
{|\mathcal G|}  \prod_{k\in \mathcal G} \big( {g_k}  {w'_k}\big)^{\frac{1}{|\mathcal G|}}=1\Big\}.\end{align*}
However, setting $W_{{0}}(v)=\prod_{k\in \mathcal G}\big(\frac{v}{g_k}\big)^{\frac{1}{|\mathcal G|}}$, we have  for every price vector $w\in \Real_{+}^d$
$$\sup_{v} \big\{W_{{0}}(v): w.v\leq\Pi_{z,\mathcal G}(w)\big\}\leq \sup_{v} \big\{W_{{0}}(v): w.v\leq\Pi_z(w)\big\}.$$
Moreover, from the Hanh-Banach extension theorem,  one can find some $\bar w \in \Real^d$ such that $\Pi_{z,g}(\bar w^\star)=\Pi_{z}(\bar w)$. In such a case:
$$\Pi_{z }(\bar w )-\bar w .z=\Pi_{z,g}({\bar w}')-{\bar w}'.z',$$
and we deduce $(d)$. $(e)$ Similarly, the result is a consequence of consequence of Proposition \ref{dualneg}.(a).
$(f)$ is the standard result established in \cite{ccf98}.  $\Box$\\

\subsection*{Numerical Example}

In the following we propose an example 
in the case of a non-parametric technology presented in example \ref{NonPar}. Let $A\subset \Real_-^m\times \Real^n$ be a finite set of observed production vectors and let $T_C=\big\{u\in \Real^d: u\leq \sum_{a\in A}t_a a, t\geq 0  \} $ be a production set satisfying an assumption of constant returns to scale. This model was proposed in \cite{ccr78}.    It follows that we have:

\begin{equation}\mathscr D_{W_{{p},g}}(z)=\sup_{t\in \Real_+^{|A|} } \Big \{ \frac{1}{{|\sss{\mathcal G}|}^{\frac{1}{p}}}\stackrel{\phi_p}{\sum\limits_{k\in
\mathcal G}}\,\frac{u_k -z_k}{g_k}: u=\sum_{a\in A}t_aa, z\leq u, t\geq 0 \Big\}.\end{equation}
Note that this formulation is related to the slack-based approach proposed by \cite{fw09} and \cite{t01}. From the proposition \ref{maxfar}, the problem can be converted to solving a linear program when $p=-\infty$ and $p=1$. If $p=\infty$, $d$ linear programs are required. When $p<1$ concave programming techniques can be used to solve it. The generalized mean directional distance function can be computed solving the following program:

\begin{align}\mathscr D_{W_{{p}},g}(z)= \max_{\delta\geq 0, t\in \Real_+^{|A|} } \notag \frac{1}{{|\sss{\mathcal G}|}^{\frac{1}{p}}}& \stackrel{\phi_p}{\sum\limits_{k\in
\mathcal G}}\,\frac{u_k-z_k}{g_k}\\ &u=\sum_{a\in A}t_aa,\\&z\leq u ,  t\geq 0.\notag\end{align}
In a more standard form, we have:

\begin{align}\mathscr D_{W_{{p}}}(z)= \max_{\delta\geq 0, t\in \Real_+^{|A|} } \notag \frac{1}{{|\sss{\mathcal G}|}^{\frac{1}{p}}}&\stackrel{\phi_p}{\sum\limits_{k\in
\mathcal G}}\, {\delta_k} \\ &z+\delta \odot g\leq \sum_{a\in A}t_a a,\\&  t\geq 0.\notag\end{align}

 {The following numerical example in Table 1 is found in \cite{fgl85}. Two inputs jointly produce a constant output level. } We use a different approach from that proposed in \cite{bck20}. In \cite{bck20}, for each parameter $p$, a specific CES-CET technology was chosen so as to reduce the problem to a linear program. Here the technology is fixed and the programs are non-linear except in the cases $p=-\infty$, $p=1$ and $p=+\infty$ (where $d=3$ linear programs are required for each firm). Note that each firm produces the same amount of output. However, since the technology is assumed to satisfy the assumption of constant returns to scale, not all firms are weakly efficient in the graph. The proposed example allows a comparison with the result proposed in \cite{bck20} with the generalized Färe-Lovell measure. 

\medskip

\begin{center}
{\scriptsize
{\begin{tabular}
{|p{0.6cm}||p{0.99cm}|p{1.2cm}|p{1.2cm}|} \hline Firms & Input 1& Input 2& Output \\ 
\hline \hline 1 & 1& 2&  2\\ 
\hline 2 & 2& 2&  2\\
\hline 3 & 2& 1&  2\\
\hline 4 & 1& 3&  2\\
\hline 5 & 1& 4&  2\\
\hline 6 & 3& 5/4& 2\\
\hline 7 & 4& 5/4& 2\\ \hline
\end{tabular}}\\ Table 1. Numerical Example}
\end{center}

The corresponding set of netput vectors is then: 
\begin{equation}A=\Big\{(-1,-2,2), (-2,-2,2), (-2,-1,2), (-1,-3,2),(-1,-4,2), (-3,-5/4,2), (-4,-5/4,2)\Big\}.\end{equation}

We consider the case where $g=1\!\!1_3$ is the free dimensional unit vector. 
\ 

\begin{center}
{\scriptsize
{\begin{tabular}
{|p{0.6cm}||p{1.6cm}|p{2.2cm}|p{2.2cm}|p{1.6cm}|p{1.5cm}|p{1.4cm}|p{1.9cm}|}
\hline Firms &$p=\frac{1}{10}$&$p=\frac{1}{2}$&$p=1$ \newline Dir. F\"are-Lovell   &$p=2$\newline Quadratic 
&$p=4$ &$p=100$& $p=+\infty$\newline Asym. Dir   \\
\hline 
\hline 1 &0.000&0.000&0.000	    &0.000       &0.000       &0.000       &0.000 \\
\hline 2 &0.292&0.296&0.333     &0.577       &0.759       &0.760       &1.000 \\
\hline 3 &0.000&0.000&0.000     &0.000       &0.000       &0.000       &0.000 \\
\hline 4 &0.015&0.123&0.333     &0.577       &0.759       &0.989       &1.000 \\
\hline 5 &0.021&0.240&0.666     &1.154       &1.519       &1.978       &2.000 \\
\hline 6 &0.297&0.333&0.416     &0.721       &0.949       &1.236       &1.250  \\
\hline 7 &0.405&0.527&0.750     &1.299       &1.709       &2.225       &2.250 \\
\hline
\end{tabular}}\\Table 2. Variation of the Generalized Mean Directional Distance function for Positive Values of $p$.}
\end{center}

\medskip

\begin{center}
{\scriptsize
{\begin{tabular}
{|p{0.6cm}||p{1.6cm}|p{2.2cm}|p{2.2cm}|p{1.6cm}|p{1.5cm}|p{1.4cm}|p{1.9cm}|}
\hline Firms &$p=-\infty$ \newline Dir. Dist. Funct.& $p=-100$& $p=-5$ & $p=-1$\newline Harmonic &$p=-\frac{1}{2}$     &$p=-\frac{1}{10}$& $p=0$ \newline Cobb-Douglas \\
\hline 
\hline 1&0.000     &0.000  &0.000  &0.000   &0.000   &0.000				&0.000 \\
\hline 2&0.285     &0.286  &0.287  &0.289   &0.290   &0.289 		    &0.292 \\
\hline 3&0.000     &0.000  &0.000  &0.000   &0.000   &0.000             &0.000 \\
\hline 4&0.000     &0.000  &0.000  &0.001   &0.001   &0.001      	    &0.000 \\
\hline 5&0.000     &0.000  &0.000  &0.001   &0.001   &0.001  		    &0.000 \\
\hline 6&0.166     &0.168  &0.183  &0.233   &0.250   & 0.226            &0.289 \\
\hline 7&0.166     &0.168  &0.183  &0.246   &0.290   &0.310  			&0.381 \\
\hline
\end{tabular}}\\Table 3. Variation of the Generalized Mean Directional Distance function for Negative Values of $p$.}
\end{center}

 {In Table 2, we consider positive values of $p$. The results are also 
compared to the Directional F\"are-Lovell distance function scores $(p=1)$ computed using the traditional linear programming  {orange}{DEA} model. In Table 2,   $\mathscr D_{W_{{p}}}(z)$ is computed for each parameter $p$ with respect to the technology $T_{A}$. One can check that when $p\rightarrow \infty$ the Asymetric Directional distance functions scores are close to the $\mathscr D_{W_{{p}}} $ score with $p=100$. This is a consequence of Proposition \ref{LimGen} and Proposition \ref{maxfar}. In addition, note that the $\mathscr D_{W_{{p}}}$ scores are non-decreasing with respect to $p$ that is a consequence of Proposition \ref{monotoneNew}. In addition, note that for each value of $p$, only firms 1 and 3 are efficient. This illustrates Proposition \ref{AXPROP}.

 {Table 3 considers negative values of the parameter $p$. The procedure is similar and one can see that when $p\rightarrow -\infty$, the efficiency scores computed for each decision making unit are close to those derived from the Directional Distance Function. When $p$ is closed to $0$ the results converges to those obtained in the multiplicative case.  These numerical results illustrate Proposition \ref{LimGen} and Proposition \ref{maxfar}. As in the case where $p$ is positive,  the $\mathscr D_{W_{{p}}} $ scores are non-decreasing with respect to $p$. For firms 4 and 5, we can note very slightly positive inefficiency scores for some values of $p$ that are negative and close to $0$. This is due to numerical reasons.  Recall that when $p<0$
 the map $\delta \mapsto \stackrel{\phi_p}{\sum_{k\in \mathcal G}}{\delta_k}$ is defined for   zero values using  a continuous extension of the generalized mean.

We consider now the case of the Generalized F\"are-Lovell measure. Let us identify $z$ with $(x,y)$ and $a$ with $(b,c)$ for each $a\in A$.  From \cite{bck20},  we have to solve the following program:
\begin{align}\mathrm{E}_{{p}}(x,y)= \max_{\lambda\geq 0, t\in \Real_+^{|A|} } \notag \frac{1}{{m}^{\frac{1}{p}}}&\stackrel{\phi_p}{\sum\limits_{k\in
\sss X}}\, {\lambda_k} \\ &\lambda \odot x\leq \sum_{(b,c)\in A}t_{(b,c)} b, \\ &y\leq \sum_{(b,c)\in A}t_{(b,c)} c,\notag \\& \notag t\geq 0, \lambda_k\leq 1, k\in X  .\notag\end{align}

\begin{center}
{\scriptsize
{\begin{tabular}
{|p{0.6cm}||p{1.6cm}|p{2.2cm}|p{2.2cm}|p{1.6cm}|p{1.5cm}|p{1.4cm}|p{1.9cm}|}
\hline Firms &$p=\frac{1}{10}$&$p=\frac{1}{2}$&$p=1$ \newline F\"are-Lovell    &$p=2$ Quadratic 
&$p=4$ &$p=100$& $p=+\infty$ \newline Farrell   \\
\hline 
\hline 1 &1.000	&1.000     &1.000		&1.000    	&1.000   	&1.000   	&1.000 \\
\hline 2 &0.711	&0.707     &0.750      	&0.750    	&0.750    	&0.750     	&0.750 \\
\hline 3 &1.000	&1.000     &1.000       &1.000      &1.000      &1.000      &1.000 \\
\hline 4 &0.818	&0.816     &0.833     	&0.849   	&0.879   	&0.993      &1.000 \\
\hline 5 &0.711	&0.728     &0.750      	&0.790   	&0.853   	&0.993      &1.000 \\
\hline 6 &0.730	&0.730     &0.733     	&0.736   	&0.742   	&0.795      &0.800  \\
\hline 7 &0.634	&0.641     &0.650      	&0.667   	&0.697   	&0.794      &0.800 \\
\hline
\end{tabular}}\\Table 4. Variation of the Generalized F\"are-Lovell measure for Positive Values of $p$.}
\end{center}

\medskip

\begin{center}
{\scriptsize
{\begin{tabular}
{|p{0.6cm}||p{1.6cm}|p{2.2cm}|p{2.2cm}|p{1.6cm}|p{1.5cm}|p{1.4cm}|p{1.9cm}|}
\hline Firms &$p=-\infty$ \newline Asym. Fare.  & $p=-100$& $p=-5$ & $p=-1$ \newline Harmonic &$p=-\frac{1}{2}$  \newline  &$p=-\frac{1}{10}$& $p=0$ \newline Cobb-Douglas \\
\hline 
\hline 1&1.000    	&1.000  &1.000  &1.000   &1.000    &1.000	&1.000 \\
\hline 2&0.500      &0.503  &0.570  &0.666   &0.686    &0.702 	&0.707 \\
\hline 3&1.000      &1.000  &1.000  &1.000   &1.000    &1.000   &1.000 \\
\hline 4&0.666    	&0.671  &0.747  &0.800   &0.808    &0.815   &0.816 \\
\hline 5&0.500     	&0.503  &0.570  &0.666   &0.686    &0.702  	&0.707 \\
\hline 6&0.583    	&0.587  &0.661  &0.727   &0.728    &0.734   &0.730 \\
\hline 7&0.437    	&0.440  &0.500  &0.608   &0.623    &0.634  	&0.632 \\
\hline
\end{tabular}}\\Table 5. Variation of the the Generalized F\"are-Lovell measure for Negative Values of $p$.}
\end{center}

 {In Table 4, paralleling the directional case, we consider positive values of $p$. We retrieve the F\"are-Lovell efficiency scores when $(p=1)$. In Table 2, one can check that when $p\rightarrow \infty$ the Farrell efficiency scores are close to the $\mathrm{E}_{{p}}$ score with $p=100$ as stated in \cite{bck20}.    These numerical results illustrate the limit properties established in \cite{bck20}. Also note that $\mathrm{E}_{{p}}$ is nonincreasing with respect to $p$ (see also \cite{bck20})}. 

 Table 5 considers the negative values of the parameter $p$. The procedure is similar and one can see that when $p\rightarrow -\infty$, the efficiency scores computed for each decision-making unit are close to those derived from the asymmetric Färe measures.

\end{document}